\documentclass{emulateapj}

\begin{document}
\title{A Hidden Population of Massive Stars with Circumstellar Shells Discovered with the Spitzer Space Telescope} 

\author{Stefanie Wachter, Jon C. Mauerhan, Schuyler D. Van Dyk, D. W. Hoard}  
\affil{Spitzer Science Center, California Institute of Technology, 
MS 220-6, Pasadena, CA 91125}
\email{wachter@ipac.caltech.edu, mauerhan@ipac.caltech.edu, vandyk@ipac.caltech.edu, hoard@ipac.caltech.edu}

\author{Stella Kafka\altaffilmark{1}}
\affil{Department of Terrestrial Magnetism. Carnegie Institution of
Washington, 5241 Broad Branch Road NW, Washington, DC 20015}
\email{skafka@dtm.ciw.edu}

\altaffiltext{1}{Visiting Astronomer, Cerro Tololo Inter-American Observatory,
which is operated by the Association of Universities for Research in 
Astronomy, Inc.\ under cooperative agreement with the National Science 
Foundation.}

\and
\author{Patrick W. Morris}
\affil{NASA Herschel Science Center, California Institute of Technology,
Pasadena, CA 91125}
\email{pmorris@ipac.caltech.edu}

\begin{abstract}

We have discovered a large number of circular and elliptical shells at 24$\mu$m around 
luminous central sources with the MIPS instrument on-board the {\sl Spitzer Space Telescope}.
Our archival follow-up effort has revealed 90\% of these circumstellar shells to be 
previously unknown.  
The majority of the shells is only visible at 24$\mu$m, but many of the central stars are detected 
at multiple wavelengths from the mid- to the near-IR regime. The general lack of optical counterparts, 
however, indicates that these sources represent a population of highly obscured objects. 
We obtained optical and near-IR spectroscopic observations of the central stars 
and find most of these objects to be massive stars. In particular, we identify a large 
population of sources that we argue 
represents a narrow evolutionary phase, closely
related or identical to the LBV stage of massive stellar evolution.  

\end{abstract}


\section{Introduction~\label{s-intro}}

The {\sl Spitzer Space Telescope} has provided us with an unprecedented
view of the mid-IR sky in terms of both resolution and sensitivity.
Invariably, the opening of every new
wavelength window carries with it the potential for the discovery of
previously unrecognized phenomena and source populations, which have gone
unnoticed at other wavelengths. 
Motivated by our discovery of an unusual well-defined elliptical shell
-- lacking a central energizing source and only visible at 24$\mu$m --
in some of the first data from the {\sl Spitzer} mission
\citep{morris06}, we executed a search for similar objects that might
shed light on its still puzzling origin and nature. 

We utilized the publicly available 24$\mu$m data products (delivery Version 3.0) of the 
MIPSGAL Legacy project \citep{carey09} obtained with the Multiband Imaging Photometer for
{\sl Spitzer} (MIPS, \citealt{rieke04}). We conducted our search by eye, inspecting all the images
of the inner 248 square degrees of the Galactic plane covered by the MIPSGAL survey.  
Although there was no prediction of whether 24$\mu$m shells would be rare or common occurrences, we 
were still surprised by our discovery of a large number of prominent, more or less ring-shaped 24$\mu$m objects: we find
over 200 such shells in the MIPSGAL survey area. 
Unlike the shell that originally motivated the search, however, the majority of the objects 
have some indication of a central source, at times quite luminous even at 24$\mu$m.
 
\begin{figure}
\epsscale{0.95}
\plotone{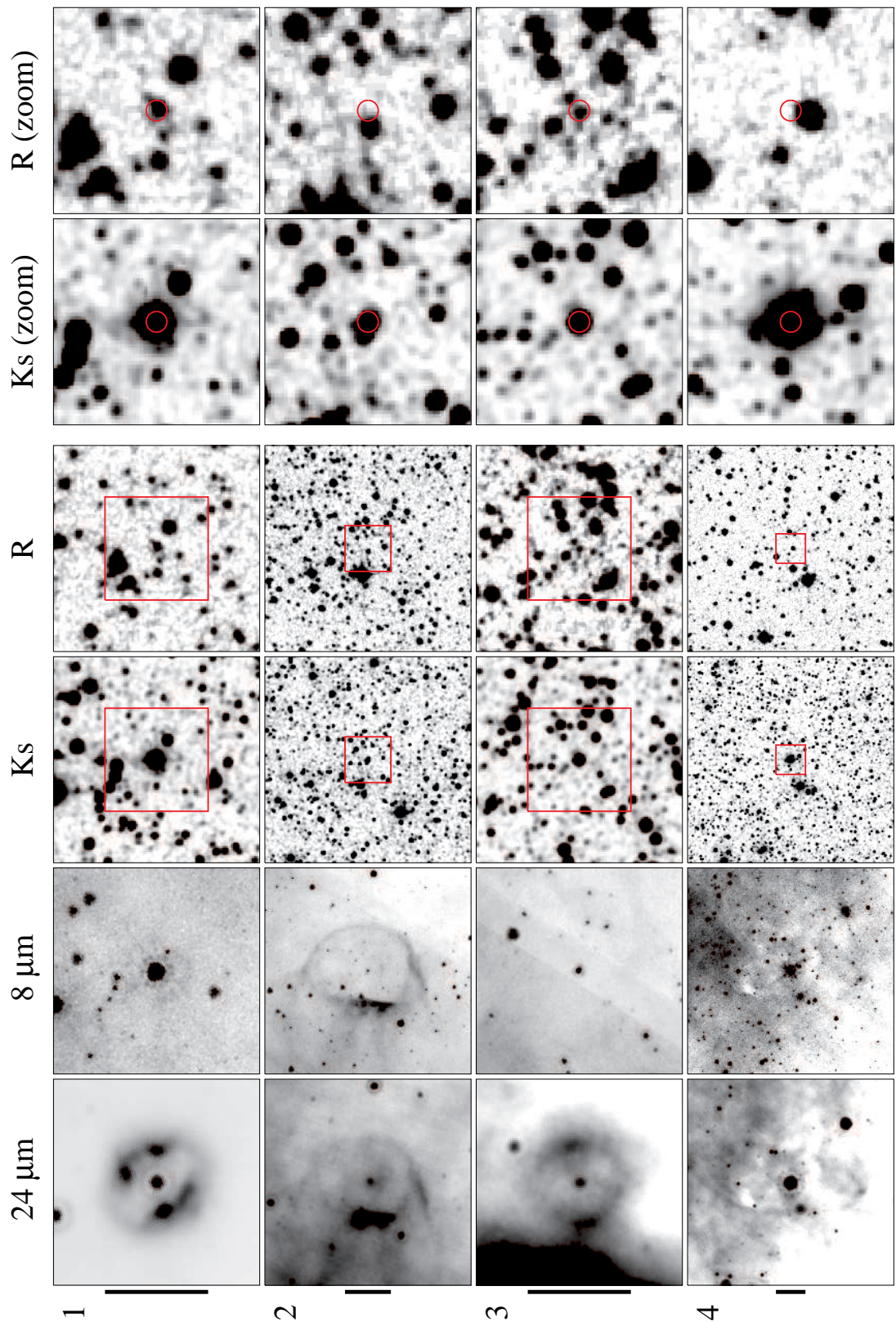}
\figcaption{MIPS 24$\mu$m images of our newly discovered circumstellar shells together with the corresponding
IRAC 8$\mu$m, 2MASS $Ks$ and DSS $R$ band images. North is up and east is to the left. A 1\arcmin\ scale bar is 
indicated on the left 
hand side of each row of panels. The assumed central source for each shell is marked. The $Ks$ and $R$ band images
are displayed at two different scales, the two rightmost panels have a FOV of 30\arcsec $\times$ 30\arcsec. 
A version of the paper including the complete (online-only) Figure 1 can be downloaded from http://web.ipac.caltech.edu/staff/wachter/bubbles/wachter\_online.pdf.  
\label{f-charts} }
\end{figure}

Follow-up investigation utilizing the complementary data from the
GLIMPSE {\sl Spitzer} Legacy
project \citep{benjamin03} which covers the same survey area with the IRAC 3.5--8.0$\mu$m bands,
as well as 2MASS and the Digitized Sky Survey (DSS), shows that about 60\% of these shells
are {\it only} detected at 24$\mu$m (see Figure~\ref{f-charts}).
This is somewhat unusual, as we generally expect a strong emission component due to polycyclic
aromatic hydrocarbons (PAHs) at 8$\mu$m if we are simply observing warm dust continuum emission.
In the case of the shell discussed by \cite{morris06}, the lack of 8$\mu$m detection is explained by a
pure line emission spectrum with a dominant [OIV] 25.89$\mu$m line. We speculate that some 
of our 24$\mu$m--only shells might exhibit similar mid-IR spectral characteristics.   

Ring-like shell morphology is observed frequently in planetary nebulae or in material associated with 
a variety of massive stars, such as Wolf-Rayet (WR) stars  
and red and blue supergiants \citep{chu91}. So far, most known circumstellar rings and shells have largely been studied using 
narrow band optical filters (e.g.\ H$\alpha$, [OIII]) and, despite being labeled as ring-shaped, 
many of these nebulae appear quite amorphous.  
For our {\sl Spitzer}--discovered shell sources, 
a SIMBAD search within 2\arcmin\ of the shell locations revealed that most (90\% of the shells, 80\% of the 
central sources) of these 
objects have not been previously studied. We present here the results of our efforts to determine
the nature of the central stars that produce the 24$\mu$m shells and to characterize
the properties of this substantial population of new sources. 
 
\section{Source Selection}

Identifying a region of 24$\mu$m emission as a more or less symmetrical shell is necessarily a 
subjective undertaking. In order to compile the most homogeneous sample possible we have selected
a subsample of our shells for dedicated follow-up observations, based on the following criteria:
\begin{itemize}
\item{The 24$\mu$m shell has a symmetric (circular or elliptical) shape with a well defined boundary.}
\item{The presence of a likely central source, detected at a minimum of two different wavelengths among 
the 24$\mu$m, 8.0$\mu$m, 3.6$\mu$m, 2MASS $Ks$ and $J$ images.}
\end{itemize}

The second criterion is particularly tricky, since we are dealing with the crowded fields in the 
Galactic plane and, hence, a relatively high a priori probability for chance superpositions. 
In order to classify a given source
as a ``likely'' central source, we first fitted each shell by eye with a circle or ellipse, then 
searched for sources within 1\arcsec\ of the geometric center of the fitted shape.  
Shells with bright 24$\mu$m central sources represent the most reliable cases, since 
the overall source 
density at this wavelength is significantly lower than in the shorter
wavelength bands. In other cases the central source identification is more tentative, and we comment 
on this for individual sources in Section~8. 
Larger shells have greater uncertainties in 
the determination of the geometric center purely from the error inherent in the fitting process.   
In addition, the literature certainly contains examples of shells 
excited by sources that are significantly offset from the geometrical center 
(e.g. \citealt{whitehead88, siebenmorgen98}).
 
Our list of targets is presented in Table~\ref{t-targets}.
For each shell we have indicated the 
2MASS designation of the likely central source, the approximate angular size of the shell, 
the presence of the central source counterpart at various wavelengths, and whether 
the 24$\mu$m shell is also detected at 8$\mu$m. We searched for known objects at the position of the 
central sources in SIMBAD, using a 2--5\arcmin\ radius depending on the size of each shell. The results
are listed in the last column of Table~\ref{t-targets}, as well as 
the results of our IR and optical spectroscopic follow-up of the central sources. 

In order to facilitate the correlation of central source properties with the morphological 
characteristics of the shells, {\sl Spitzer} 24 and 8$\mu$m, 2MASS $Ks$ and DSS $R$ band images of the shells are displayed in 
Figure~\ref{f-charts} (in its entirety available only in electronic format). The assumed central source is 
marked.

\section{Observations of the Central Sources}

While some of the central sources from Table~\ref{t-targets} can be matched to objects 
in SIMBAD, the vast majority has not been observed before.  
It is impossible to determine the nature of the central sources based solely on 
archival photometric data, since the extinction -- both intrinsic to
the source and between us and the source -- and the distance to the shells are unknown. Hence we 
obtained optical and IR spectroscopy of the central sources of 
these newly discovered shells to identify their nature. The spectroscopic data obtained is indicated 
by ``IR'' for infrared and ``OPT'' for optical in Table~\ref{t-targets}, respectively.   

\subsection{Near-Infrared Spectroscopy}

Near-Infrared spectroscopy of a number of the central stars of our shells was obtained on 
2009 June 23--25 UT with the 
Ohio State Infrared Imager and Spectrograph (OSIRIS; \citealt{depoy93}) on the
Southern Observatory for Astrophysical Research (SOAR) 4.1m telescope. Our low-resolution 
spectra (R$\sim$1200) were obtained using the 1\arcsec\
cross--dispersed slit and the f3 camera which covers the $J$, $H$ and $K$ bands simultaneously. 
Spectra were acquired in an ``ABBA'' nodding sequence for sky subtraction. 
We utilized sky spectra for wavelength calibration.  
Data reduction and spectral extraction and calibration was performed in the standard 
manner using IRAF. 
For telluric correction, spectra of bright A0V standard stars ($Ks\approx6$--7 mag) 
were obtained.  
The IDL program {\it xtellcor}
\citep{vac03} was used, which removes model H {\sc i} absorption 
lines from the A0{\sc V} spectra before application to the science data.
Observing conditions were photometric on the first night, but strongly affected by 
variable clouds during the following two nights, which limited the observations to 
the brighter targets in our source list. Note that no $J$ band spectra could be 
extracted for stars \#4, 5, 6a, 6b, 8a, 14, 20, 22, 29, and 44.  

The IR spectrum for star \#52 was obtained with the Palomar Hale 200" telescope and the 
TripleSpec (TSPEC) near-infrared spectrograph on 2009 July 14 UT.  TSPEC provides
simultaneous coverage of the $J$, $H$, and $K$ bands ($\lambda=1.0$--2.4 {\micron}) and 
produces a moderate resolution spectrum ($R\approx2500$--2700)
through a $1\arcsec \times 30\arcsec$ slit. Spectra were obtained utilizing an ABBA 
telescope nodding sequence for sky subtraction and bad-pixel suppression.
Wavelength calibration was performed using the OH emission lines in the sky spectra. 
The A0V star HR 6958 was observed as a standard, and a telluric spectrum
was applied to the science data using the program {\it xtellcor}.

\subsection{Optical Spectroscopy}

For the limited number of sources with optical counterparts, optical spectroscopy 
was obtained with the 200" Hale telescope at Palomar Observatory 
on 2008 September 3 and 4 UT
using the Double Spectrograph. This low to medium resolution grating spectrograph 
uses a dichroic to split the light into separate red and blue channels which are observed 
simultaneously. The blue side CCD has 15$\mu$m pixels and a scale of 0.389\arcsec pixel$^{-1}$, the
red side CCD 24$\mu$m pixels with a scale of 0.468\arcsec pixel$^{-1}$. We used the 316 lines/mm 
grating in first order, dichroic D55 and a slit width of 1\arcsec. The resulting spectra cover
4000--5600\AA\ and 5800-8300\AA\ with a
dispersion of 2.0 and 2.4\AA pixel$^{-1}$ on the blue and red side respectively, and have a resolution 
of 5--7\AA. For wavelength calibration, FeAr (on the blue side) and HeNeAr (on the red side) 
arc lamp spectra were obtained at each new telescope position. 
Bias and flat-field corrections to the raw spectrum images were performed with IRAF.
Various routines within the IRAF packages {\it twodspec} and {\it onedspec} were used to extract and 
wavelength/flux
calibrate the spectra. Both nights were photometric with variable seeing.  
Observations of spectrophotometric standard stars were carried out, however, these were 
mainly used to remove the instrumental response function. 
A signal-to-noise ratio of 40--100 was achieved in the final spectra. 

\section{Spectral Classification}

\subsection{IR Spectra}

At first glance, our central source IR spectra can be roughly divided into two main groups: those that 
show emission lines or early spectral type features and those that exhibit absorption features
indicative of late type stars.  

\subsubsection{Late Type Stars~\label{s-late}}

\begin{figure}[htb]
\epsscale{0.75}
\plotone{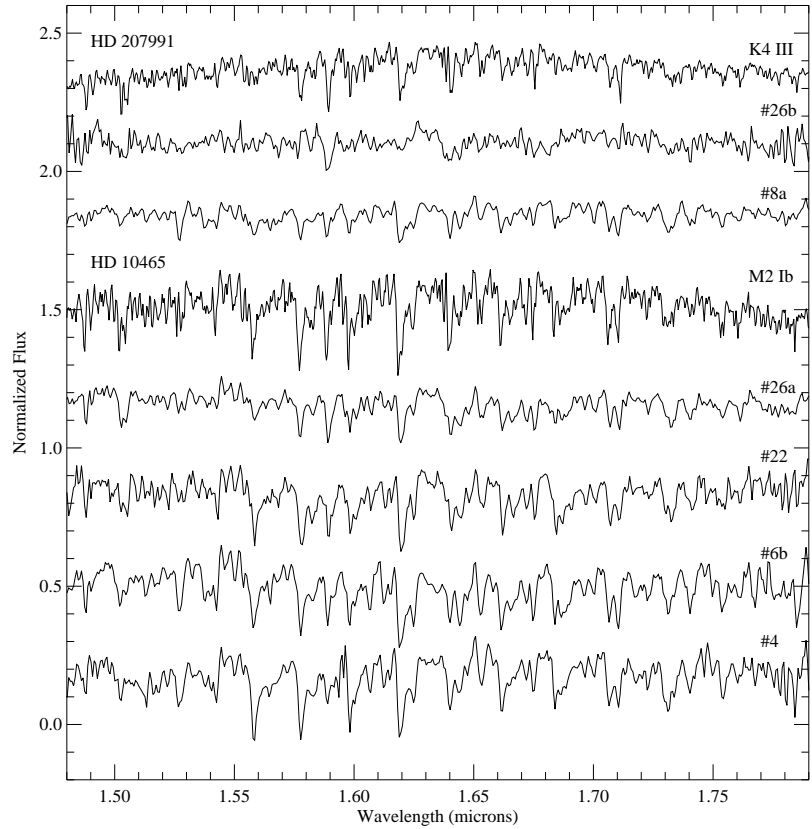}
\plotone{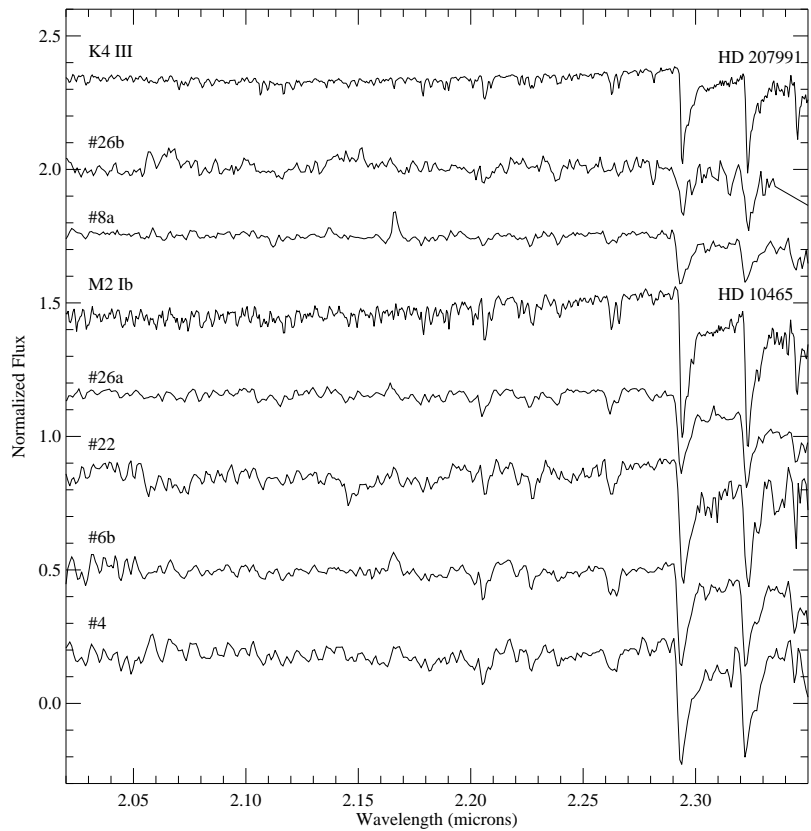}
\figcaption{$H$ and $K$ band spectra of a sample of shell central sources indicating a late type star. 
The spectra have been normalized and offset from each other for display 
purposes. For comparison the spectra of HD 207991 (K4 III) and HD 10465 (M2Ib) obtained
from the IRTF spectral library \citep{rayner09} are also shown.  
\label{f-late} }
\end{figure}

For the late type stars, we determined the spectral type and luminosity class based on the 
strength of the CO bandhead at 2.29$\mu$m, following the methodology of \cite{figer06} and 
\cite{davies07}. Both ways provided consistent results. We also compared the appearance of the 
spectra to those from the IRTF spectral 
atlas \citep{rayner09}. In addition, we performed the same analysis for selected spectra from the atlas
(of stars with known spectral type and luminosity class) 
as an independent check for the validity of our measurements. We estimate that our derived spectral types
are accurate to within $\pm 1$ subtype. The results are listed in Table~\ref{t-size} and a selection 
of our late type spectra are displayed in Figure~\ref{f-late}. Note that several sources show 
Br$\gamma$ in emission. 

We caution that veiling of the absorption features by a hot dust continuum 
could affect our spectral type determination. This would lead to an underestimate of 
our measured equivalent width, and hence the true spectral types of our sources would be later than those
we derived. In order to investigate the presence of circumstellar dust, we plotted the 2MASS colors of the stars 
from Table~\ref{t-size} in a near-IR color-color diagram (not shown).
Typically, IR excess due to circumstellar material will cause a shift to redder $H-Ks$ colors relative to 
the unreddened stellar loci. We find that all of our stars fall on the red (super)giant reddening line within 
the photometric uncertainties. The most deviant point corresponds to star \#23b, which is unlikely to produce 
the associated shell as discussed below. 

As indicated in Table~\ref{t-size}, each CO bandhead equivalent width measurement corresponds to a solution for both   
giants and supergiants, and we are unable to break the degeneracy solely based on our spectra. 
Hence we investigated whether photometric information could provide additional constraints 
to distinguish between these possibilities. For each source, we first calculated the total 
(interstellar and circumstellar) A$_V$ by  
comparing its 2MASS photometry (provided in Table~\ref{t-shellphot})
with the unreddened $J-K$ colors of the appropriate spectral type from \cite{cox00}, utilizing 
the reddening law by \cite{card89}. 
Finally, we derived the distances and the physical sizes of the surrounding shells 
for the two possible luminosity classes, respectively. 
These results are also listed in Table~\ref{t-size}. 

For all stars, the derived A$_V$ exceeds the 
average 1 mag kpc$^{-1}$ value by several factors, implying that the line of sight to these sources intersects
localized areas of high extinction. Minimizing this excess extinction would indicate the supergiant case for 
all sources. However, for  
star \#26b we favor the giant scenario, 
since the distance of $>$23 kpc for the supergiant case results in an average extinction of only 0.4 mag kpc$^{-1}$
and the largest physical shell size of the sample. 
We derive and list these parameters for all of the sources for completeness, however, 
stars \#8b, \#23b, and \#28b are probably not producing their respective observed shells.  These particular stars are 
members of {\it apparent} doubles, where the center of the fitted shell geometry coincides with the 
other component of the double. While we cannot exclude the possibility that the off--center source is
responsible for the shell, it appears more likely that the central source is producing the shell, 
in particular in the case of \#23b and \#28b, where the central source was found
to be a hot star (see the following section). 

Most of our knowledge about large scale circumstellar shells around red giants and supergiants 
dates back to the IRAS 60$\mu$m surveys of \cite{stencel89} and \cite{young93}. \cite{young93} 
surveyed 512 red giant stars, 15\% of which were found to have shells with radii ranging from 
0.1--4.6 pc. The average shell radius was 0.74 pc, with radii of $>2$ pc being quite rare. \cite{stencel89}
conducted a similar survey for 111 red supergiants, 25\% of which showed evidence for shells with 
radii between 0.1--4 pc (most clustered between 0.1--1.5 pc). 
Our shells have very similar size characteristics, which is somewhat surprising given that, a) our 
shells were observed at 24$\mu$m and one might expect size differences between the shells at 
various wavelengths, and, b) the resolution of the IRAS 60$\mu$m images was quite limiting, such that
these surveys were biased towards detecting only the largest shells of any underlying distribution.

\subsubsection{Emission Line / Early Type Stars}

Most of our IR spectra with emission lines resemble those of the so-called ``transition objects'', massive 
stars of type Of, WNL, Be, B[e] and Luminous Blue Variable (LBV), thought to span the evolutionary 
stages between O and Wolf-Rayet (WR) stars \citep{morris96}. 
The classification of these stars based on IR spectra is notoriously difficult.
We find that our spectra separate fairly cleanly into two distinct groups simply based on the 
appearance of their spectral features. When considering commonalities, we give less weight to the 
appearance of the He {\sc i} 2.058$\mu$m line, since it is highly variable due to its susceptibility to 
optical depth effects. 

\begin{figure}[htb]
\epsscale{0.55}
\plotone{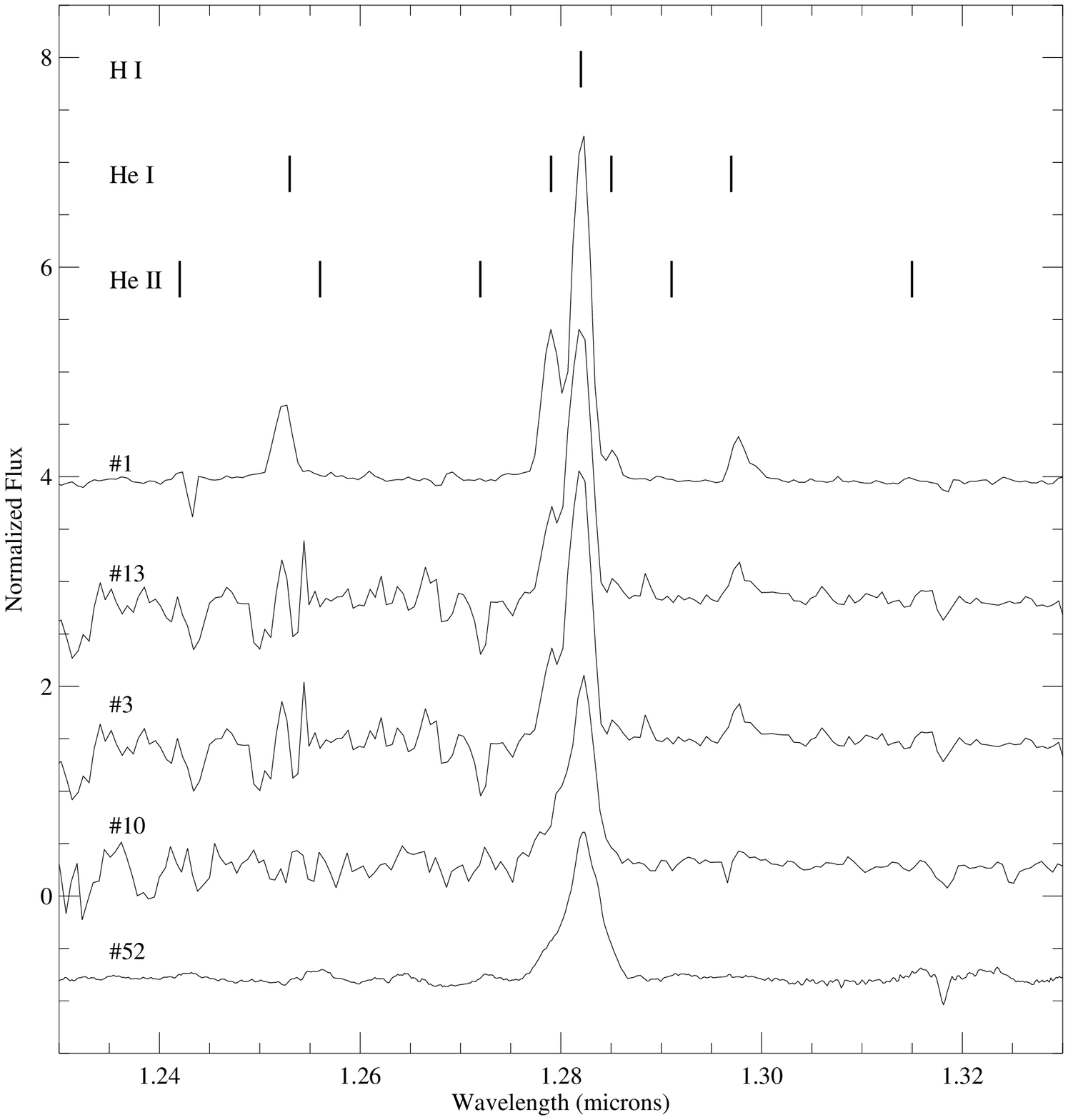}
\plotone{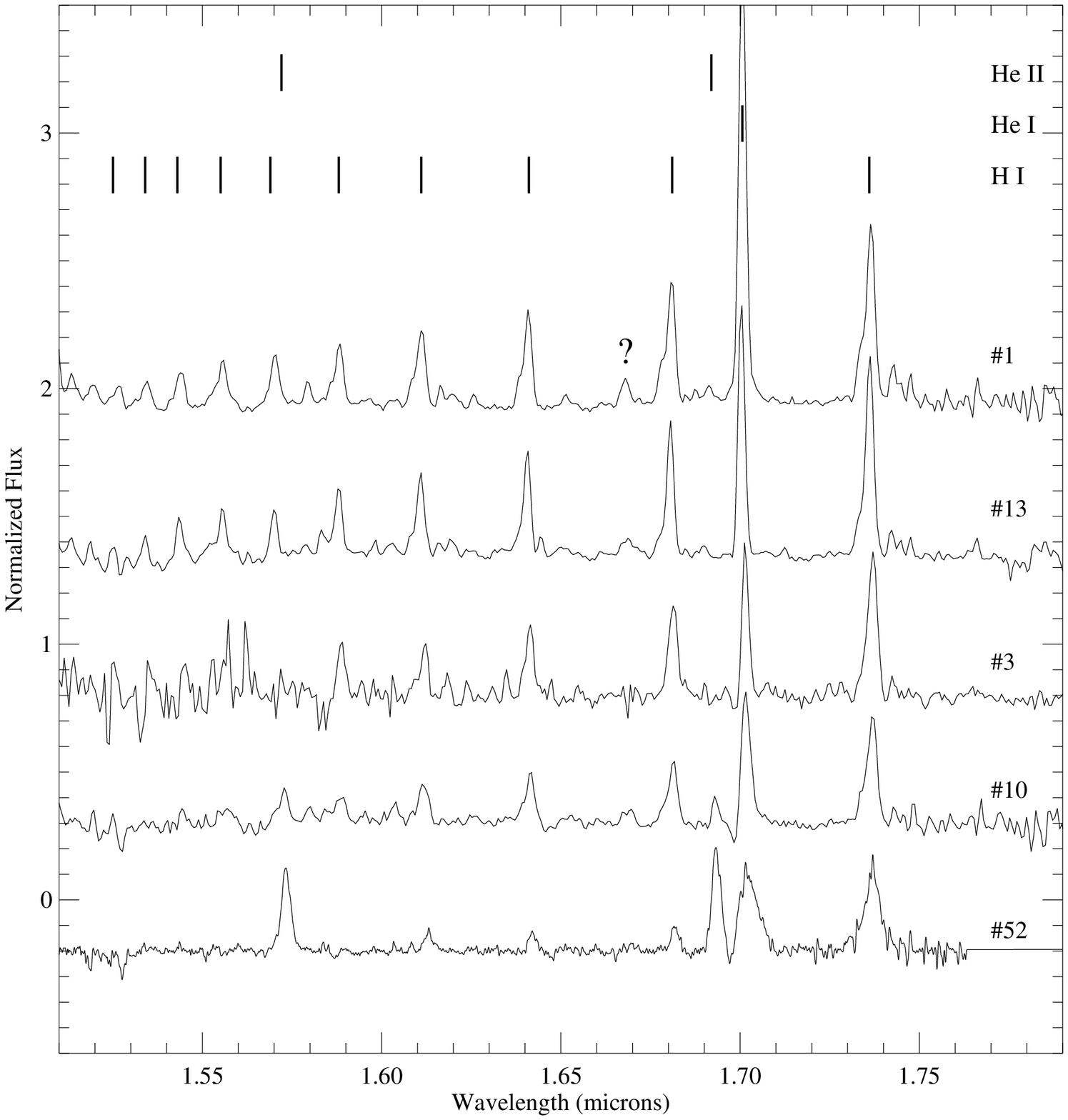}
\plotone{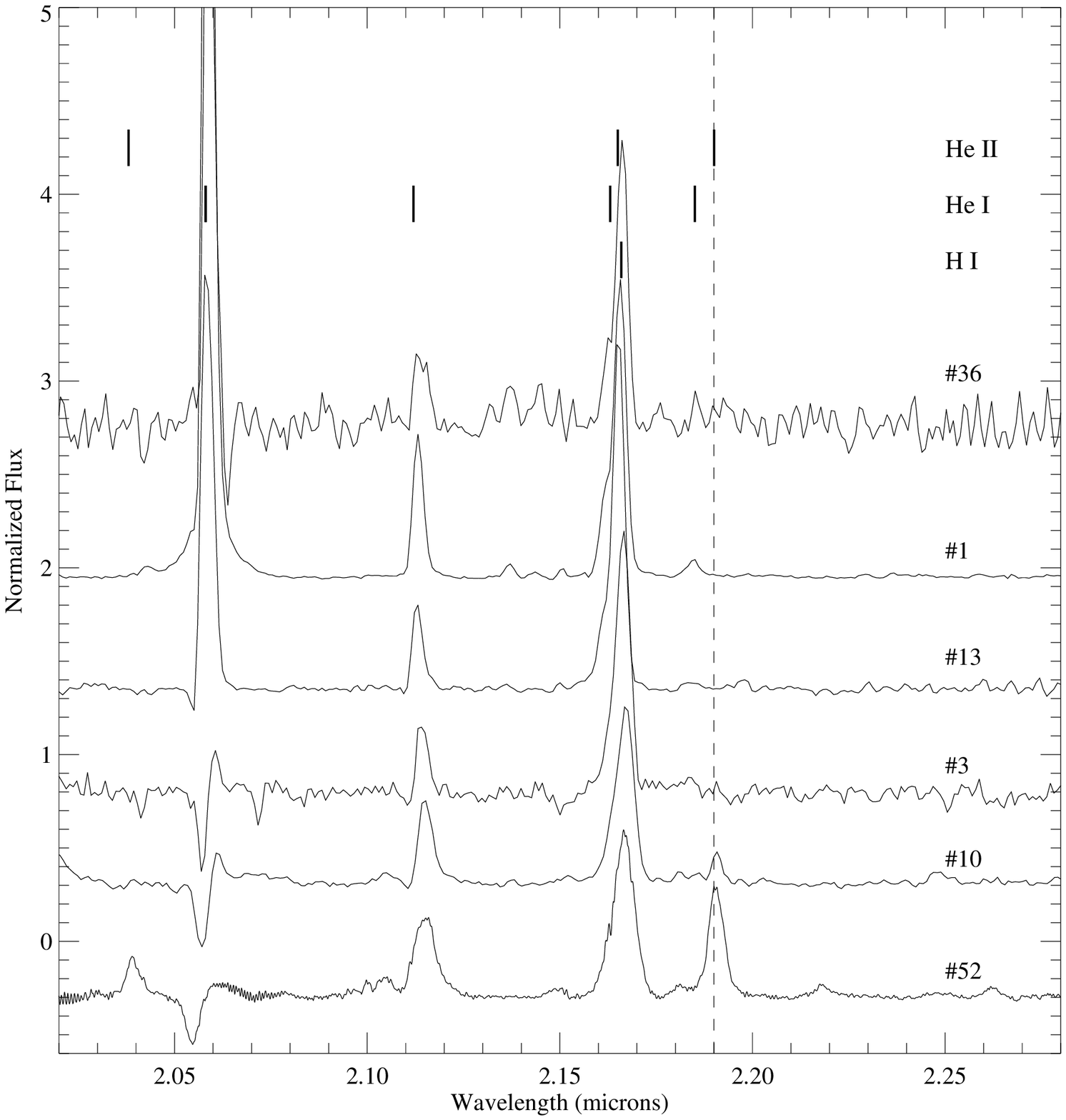}
\figcaption{$J$, $H$, and $K$ band spectra of shell central sources characterized by H  {\sc i} and He  {\sc i} (1.70 ad 2.112 $\mu$m)
emission (group 1). Star \#10 is a newly discovered WN9h star, star \#52 a newly discovered WN7 star.
The spectra have been normalized and offset from each other for display 
purposes. 
\label{f-1like} }
\end{figure}

\begin{figure}[htb]
\epsscale{0.55}
\plotone{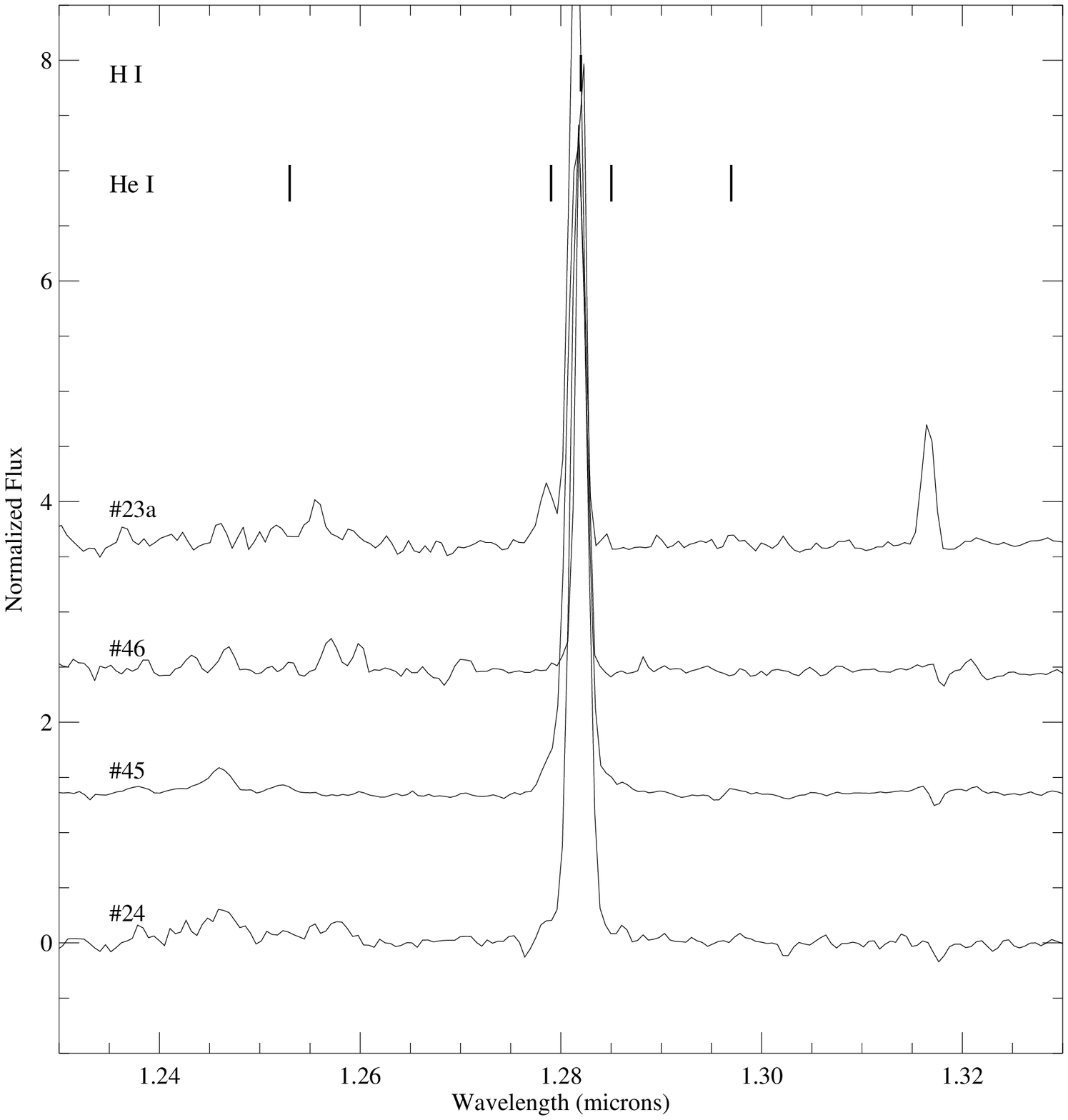}
\plotone{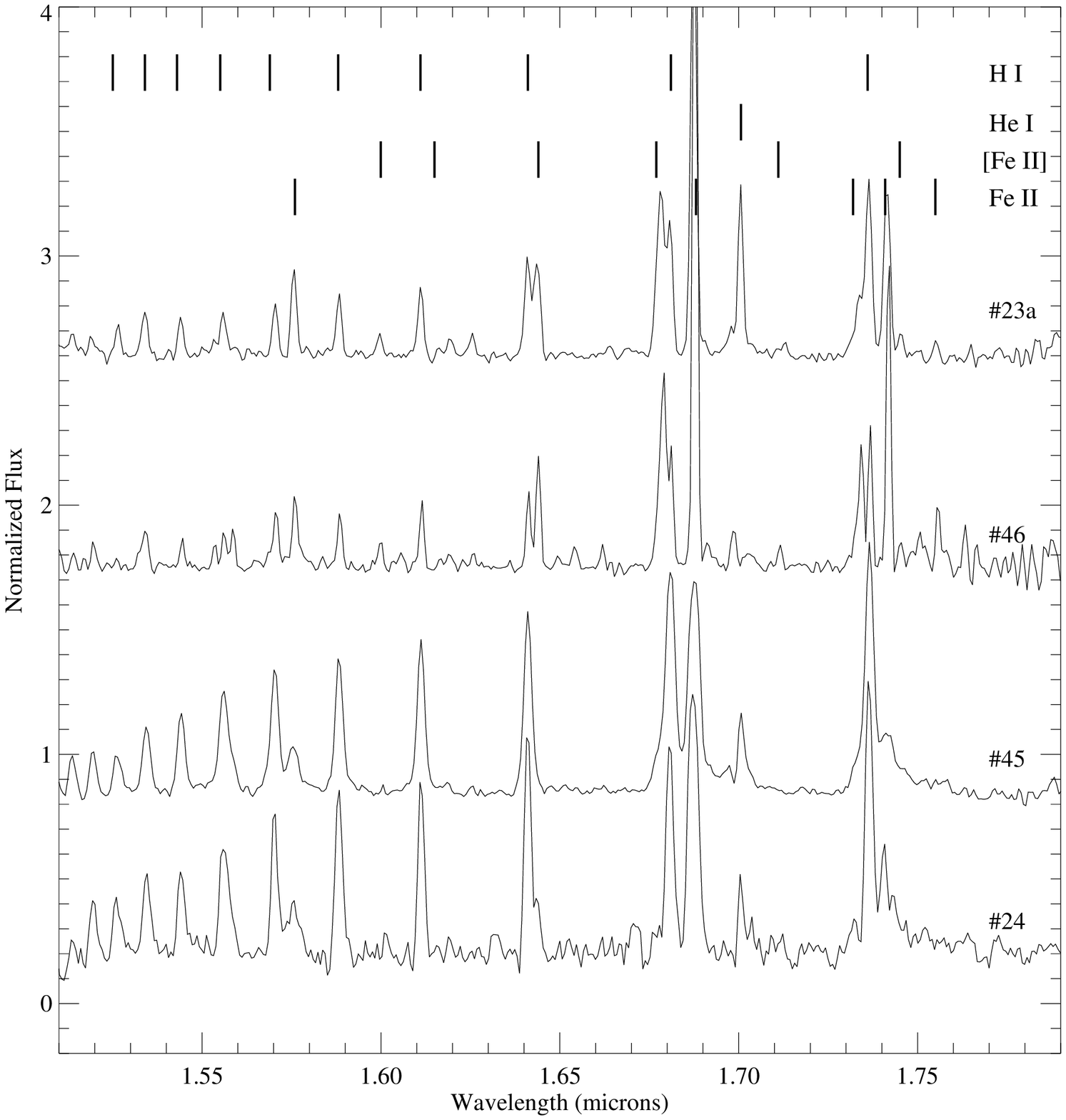}
\plotone{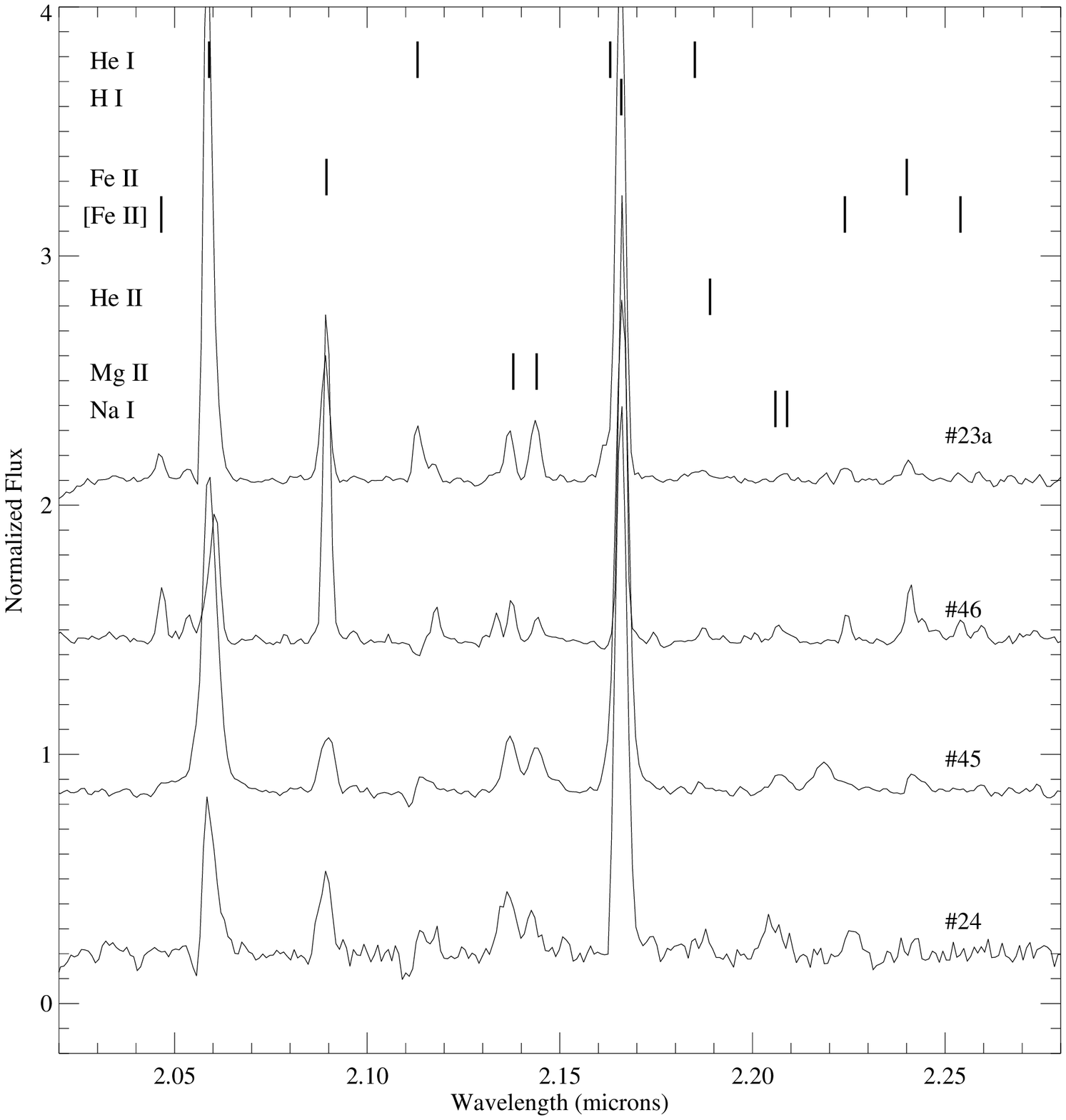}
\figcaption{$J$, $H$ and $K$ band spectra of shell central sources characterized by weak or absent
He {\sc i} (1.70, 2.112 $\mu$m) features and strong H {\sc i}, Fe {\sc ii} and
Mg {\sc ii} emission (group 2A). These sources strongly resemble the stars classified as LBVs and
Be/B[e] stars by \cite{morris96}.
The spectra have been normalized and offset from each other for display
purposes.
\label{f-46like} }
\end{figure}

\begin{figure}[htb]
\epsscale{0.55}
\plotone{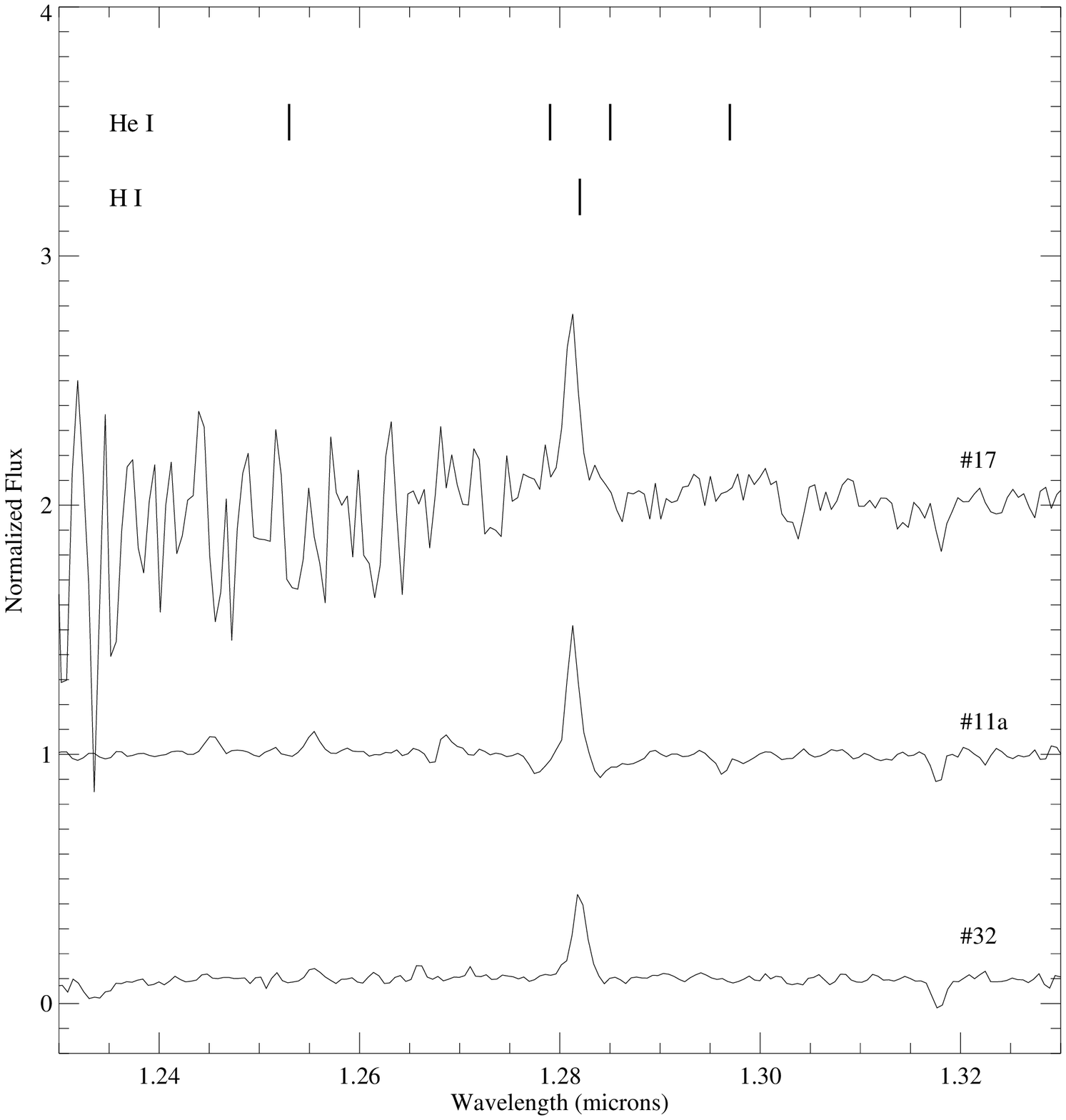}
\plotone{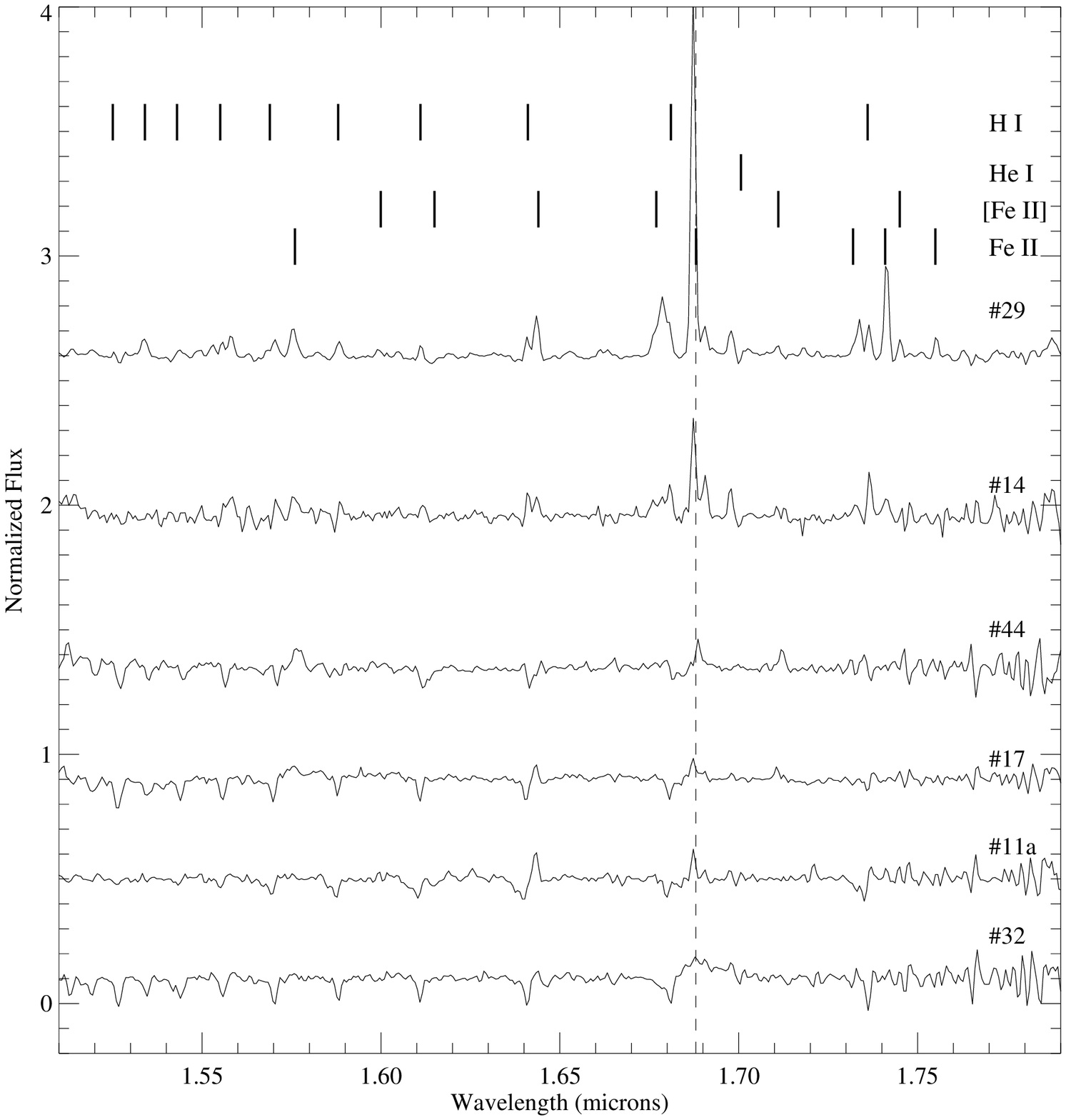}
\plotone{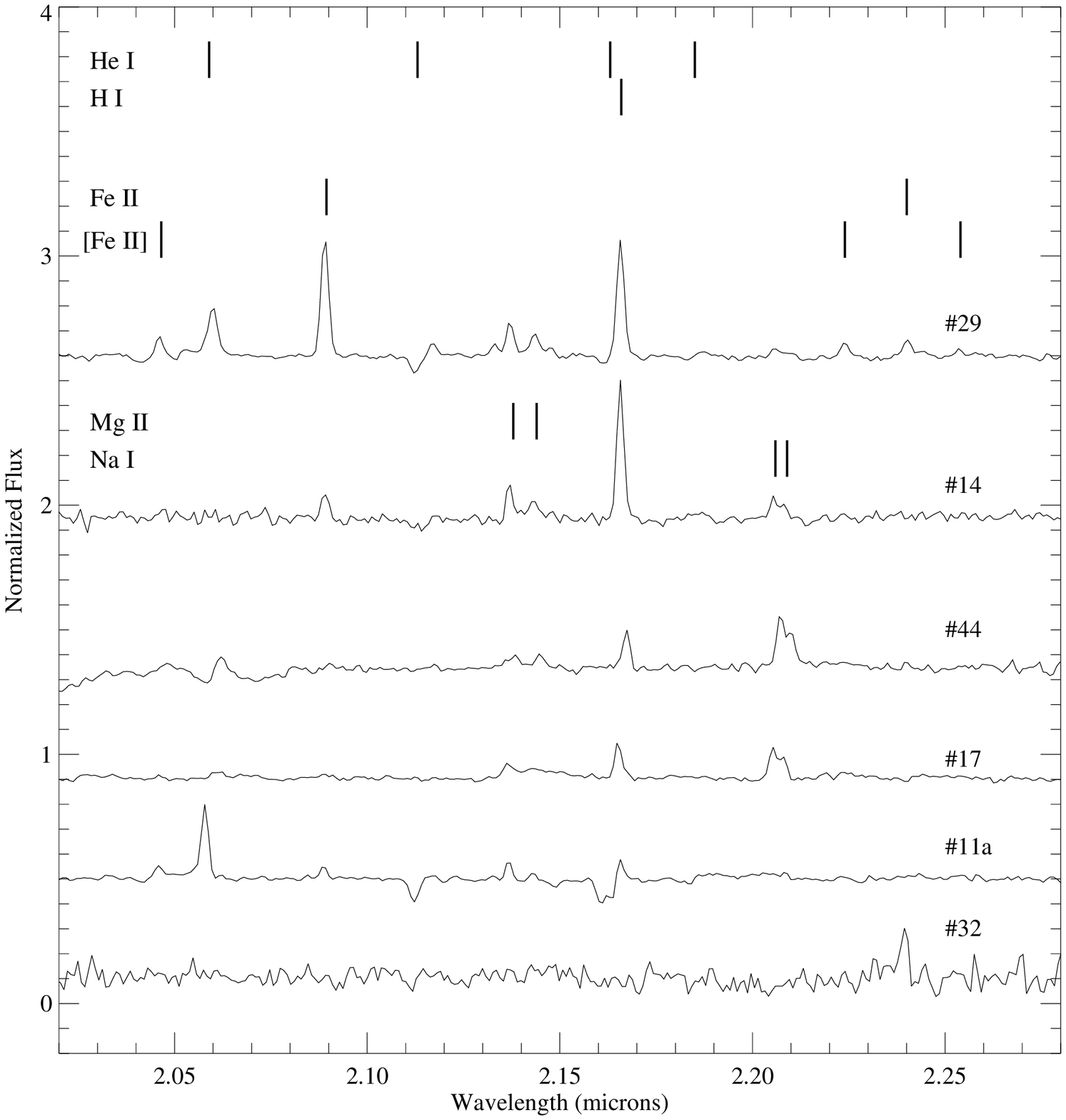}
\figcaption{$J$, $H$ and $K$ band spectra of shell central sources characterized by weak or absent 
He {\sc i} (1.70, 2.112 $\mu$m) features and strong Fe {\sc ii} or 
Mg {\sc ii} emission (group 2B). The spectra are similar to those of group 2A (Figure~\ref{f-46like}) 
but exhibit weaker H {\sc i} 
features. Note that we could not extract $J$ band spectra for all of these sources.    
The spectra have been normalized and offset from each other for display 
purposes. 
\label{f-14like} }
\end{figure}

The first group is comprised of stars 
\#1, 3, 10, 13, 36, and 52. For star \#36 we only have a $K$ band spectrum, $J$ and $H$ 
were too noisy to reliably identify features. All stars exhibit strong He {\sc i} 1.70 and 2.112$\mu$m,  
as well as prominent H {\sc i} emission in all three bands (Figure~\ref{f-1like}). For all figures, the 
line identifications have been compiled from \cite{morris96}, \cite{figer97}, and \cite{crowther06}. 
We have divided the second group into two subgroups, distinguished by the strength of the 
H {\sc i} features. The spectra in group 2A (stars \#23a, 24, 45, and 46) are still dominated by  
H {\sc i} emission of similar strength as group 1 in the $J$, $H$, and $K$ band, but the striking 
difference with the spectra of group 1
lies in the strong Fe {\sc ii} and Mg {\sc ii} emission and weak or absent He {\sc i} (1.70, 2.112$\mu$m) features (Figure~\ref{f-46like}).
In addition, [Fe {\sc ii}] is detected in all the stars except \#45. 
Group 2B (stars \#11a, 14, 17, 29, 32, 44) is characterized by much weaker H {\sc i} emission (or even weak H {\sc i} absorption)  
compared to the first two groups. However, similar to the stars from group 2A, the spectra exhibit
Fe {\sc ii} or Mg {\sc ii} emission lines (Figure~\ref{f-14like}). We regard group 2B as an extension of group 2A towards 
weaker H {\sc i} features. In particular, star \#29 could have been
assigned to either group 2A or 2B, and emphasizes the connection between the two groups. 
Star \#32 is included in group 2B based on the similarities of its spectral features in the $H$ band and 
the presence of Pa$\beta$ emission 
in the $J$ band; however, it has no obvious emission lines of Fe {\sc ii}, and the $K$ band is featureless. In SIMBAD, the
source is classified as M2: by \cite{raharto84}, but an M spectral type can be ruled out based on our IR spectrum, given the 
H {\sc i} absorption spectrum in the $H$ band and the lack of CO absorption features in the $K$ band. 

\begin{figure}[htb]
\epsscale{0.55}
\plotone{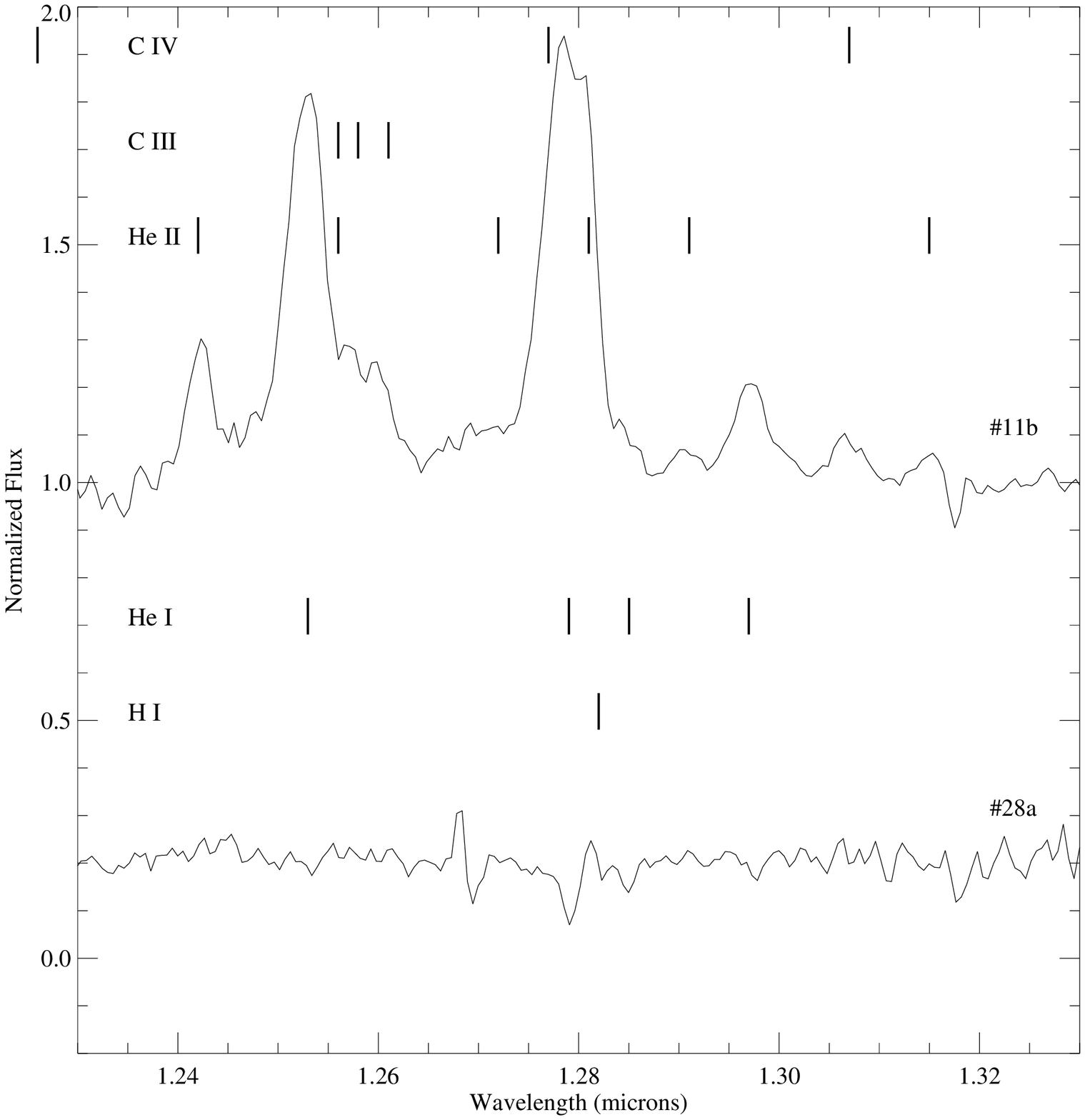}
\plotone{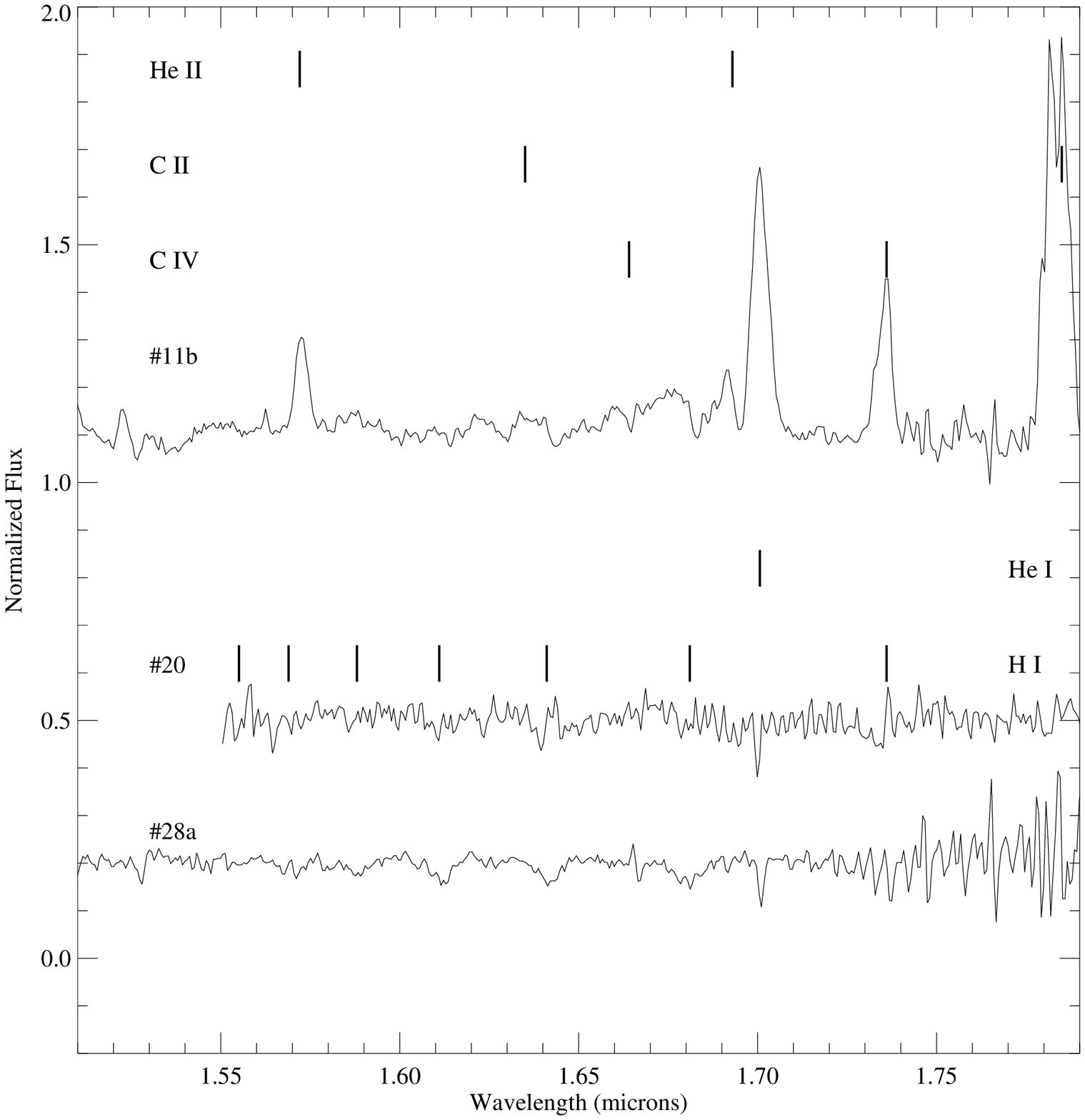}
\plotone{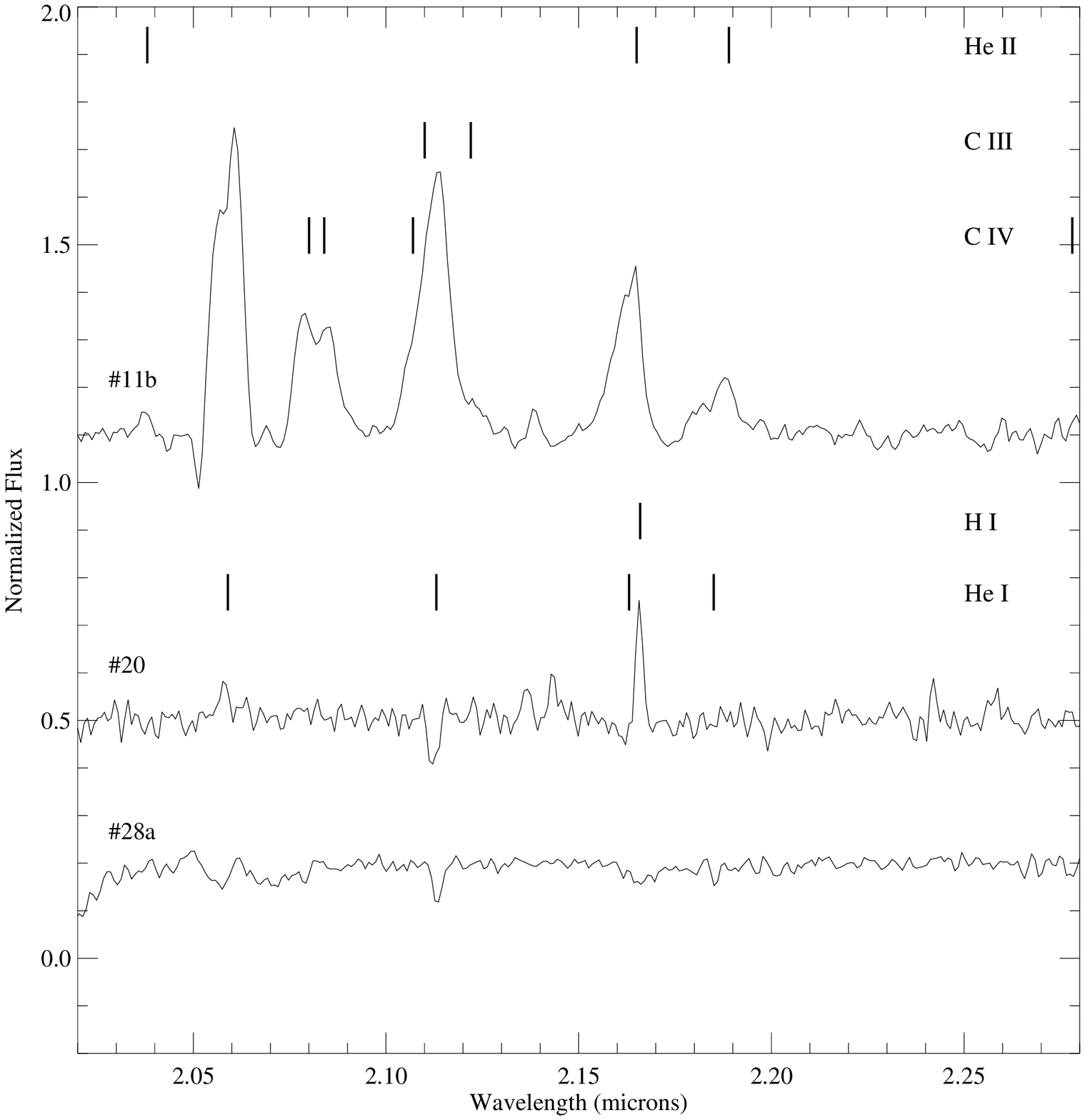}
\figcaption{Spectra of the newly discovered WC9 (\#11b) star as well as the two spectra that did not fit 
in the classification scheme of the sources in groups 1 and 2. The spectra have been normalized 
and offset from each other for display
purposes.
\label{f-other} }
\end{figure}

We inspected all of our spectra for the presence of the He {\sc ii} 2.189$\mu$m line, since this is 
one of the key lines in identifying WR and O type stars. Only star \#10, \#52, and \#11b (Figure~\ref{f-other})
exhibit convincing He {\sc ii} 2.189$\mu$m emission. Here, care must be taken to not mistake He {\sc i} 2.185$\mu$m emission
for that of He {\sc ii} 2.189$\mu$m. Our spectra have sufficient resolution to clearly distinguish between the two lines, 
e.g., star \#1 shows He {\sc i} 2.185 and not He {\sc ii} 2.189$\mu$m emission. 
The ratios of the equivalent widths (EW) of certain emission lines in the $K$-band are commonly used as a
WR subtype diagnostic (e.g., see \citealt{figer97}, \citealt{crowther06}).
For star \#52 and \#10 we measure EW($\lambda$2.189 {\micron})/EW($\lambda$2.165 {\micron})= 0.5 and $< 0.1$, 
respectively, which correspond to WR subtypes of WN7 and WN9h. We also have optical spectra of star \#52 
which independently confirm its WN7 classification based on a different set of spectral criteria (see 
Section~\ref{s-opt}). 
For star \#11b we find EW (C{\sc iv} 2.076) / EW (C{\sc iii} 2.110) $< 1$, consistent with a WC9 subtype. 

Unfortunately, no classification scheme based on the EW measurement of features exists for the
remaining spectra, and we have to rely on comparison to the various spectral atlases, e.g., \cite{hanson96} and \cite{morris96},
to classify these sources. The stars from our group 2 strongly resemble the spectra of LBV and Be/B[e] stars
from \cite{morris96} shown in their Figures 5, 6, and 7. The exception in that sample is the LBV He 3-519,
which does not exhibit Fe {\sc ii} emission, but instead looks like the spectra of stars in our group 1. 
It is interesting to note that 
[Fe {\sc ii}] 1.644$\mu$m emission has been found to be a common property of the shell emission among LBVs with 
prominent nebulae, but 
has rarely been seen in the spectra of their central sources \citep{smith02}. In contrast, we observe 
[Fe {\sc ii}] emission in 
about half the sources of group 2. 
Based on these characteristics, we tentatively classify the stars in our group 2 as Be/B[e]/LBV candidates. 
This is supported by the fact that star \#29, \#44, and \#45 are identified as LBVs or LBV candidates in the 
literature (for more details see Section~\ref{s-lbv}).   

Among the stars of group 1, we have already classified source \#52 as WN7 and \#10 as a WN9h star. The $K$ band spectra of the 
remaining four stars in group 1 (\#1, 3, 13, 36) closely match the appearance of the WN9 source, except for the 
presence of the He {\sc ii} 2.189$\mu$m line. Several studies have classified similar--looking spectra as WN9, despite the 
lack of He {\sc ii} 2.189$\mu$m emission. For example, \cite{figer99} show several spectra of Quintuplet cluster stars 
that are virtually ``clones'' of our group 1 spectra, and none exhibit evidence for features at 2.189$\mu$m.
\cite{shara09} display a $K$ band spectrum of the known WR 102d, classified as WN9, that equally lacks 
He {\sc ii} 2.189$\mu$m. 
Finally, our spectra at both $H$ and $K$ band resemble the Ofpe/WN9 stars of \cite{morris96} (their Figures 2 and 3).
Only some of those stars exhibit He {\sc ii} 2.189$\mu$m. However, the Of spectra of \cite{hanson96} demonstrate that 
He {\sc ii} 2.189$\mu$m is always present for spectral types earlier than O9. Incidentally, the Oe spectra of 
\cite{hanson96} are very similar in  
appearance to our group 1 stars and do not show He {\sc ii} 2.189$\mu$m. While the exact classification of these types of spectra
is clearly still an open question, they undoubtedly indicate massive stars most likely on their way to 
a WR phase. For the purposes of our study, we will label these stars as Oe/WN9. 

The final two spectra remaining to be classified are shown in Figure~\ref{f-other}, together with our newly discovered
WC9 star. Star \#20 exhibits Br$\gamma$ emission, He {\sc i} 2.112$\mu$m in absorption, as well as possible
Mg {\sc ii} emission in the $K$ band and He {\sc i} 1.70$\mu$m absorption and a hint of the H {\sc i} series in 
absorption. Based on the absence of obvious He {\sc ii} features, we classify star \#20 as type Be.  
Source \#28a also shows
He {\sc i} 1.70 and 2.112$\mu$m (and possibly 2.185$\mu$m) in absorption and shallow absorption of H {\sc i}.  
Based on the $K$ band spectrum and comparison to \cite{hanson96}, the spectrum of star \#28a most closely
resembles that of HD 168021 (B0 Ib), HD 213087 (B0.5 Ib) and HD 226868 (HMXB, O9.7 Iab) in the 
relative strength of the features at Br $\gamma$ and 2.112$\mu$m. However, in the $H$ band the shallow and 
broad H {\sc i} features indicate luminosity class V and, together with the strength of He {\sc i} 1.70$\mu$m line, 
a spectral type of O9--B0.5 \citep{hanson98}. We classify star \#28a as OB, noting that the luminosity class issue still 
remains to be resolved.     
The final spectral classification for 
each source is listed in Table~\ref{t-targets}.

\subsubsection{Known LBVs~\label{s-lbv}}

Sources \#29 (WRAY 17-96 = Hen 3-1453), \#44 (G024.73+00.69 = V481 Sct), and
\#45 (G026.47+00.02) are identified
in SIMBAD as a LBV candidate \citep{egan02}, confirmed LBV \citep{clark05}, and LBV candidate \citep{clark03}, respectively. 
All have 
been found to be surrounded by ring nebulae in the mid-IR with ISO \citep{egan02, clark03}. 

LBVs are generally defined as a class of massive, unstable stars that populate the upper left corner of
the HR diagram (for a detailed review see \citealt{humphreys94}). These high luminosity stars are 
thought to be a short--lived phase before the WR stage, where 
a massive star sheds a significant amount of its mass. Consequently, LBVs  
are characterized by high mass loss rates,  
significant photometric and spectral variability and occasional mass eruptions, 
the most famous 
example being the historic 10-magnitude brightening of $\eta$ Carinae in the 1840s. These outbursts are 
extremely energetic; they have been linked to the discovery of rapidly expanding nebulae surrounding 
LBVs and, in several cases, have previously been mistaken for supernovae in external galaxies 
\citep{vandyk00}. More common than such giant eruptions, however, is the presence of a circumstellar nebula or shell, 
indicating more modest, past mass ejection episodes. Out of the 12 confirmed LBVs, seven are associated 
with circumstellar ejecta, 
as well as 13 of the 23 LBV candidates listed by \cite{clark05}. The underlying physical cause that triggers the LBV 
instability is not yet understood, the current consensus invokes instabilities in the outer 
layers of the stars as they approach the Eddington luminosity during their evolution.    

Our newly acquired spectra of the known LBVs allow us to search for the expected spectral variations in comparison to the 
spectra previously published in the literature.  
LBVs in quiescence (at visual minimum) typically exhibit spectral types of Be or B[e], rich with emission lines of hydrogen,
helium, iron, and sodium. In many LBVs, the Fe {\sc ii} and [Fe {\sc ii}] emission lines are most pronounced during
the quiescent state. During an outburst, the spectra may morph into those of much cooler type A/F supergiants as 
a pseudo-photosphere is formed in the optically thick expanding stellar wind. 
In one particularly remarkable display, the quintessential LBV S Doradus changed spectral type from B2e to F
sometime between 1996 and 1999 \citep{massey00}.

Our $K$ band 
spectrum of WRAY 17-96 does not reveal any significant changes compared to the one presented in \cite{egan02}, except
in the highly variable He {\sc i} 2.058$\mu$m line. 
To our knowledge our $H$ band spectrum is the first one published of this source. 
Our $K$ band spectrum of G026.47+00.02 appears identical to that of \cite{clark03}. Our $H$ band spectrum is of
higher resolution and shows additional lines not resolved in the one by \cite{clark03}.

For G024.73+00.69, our spectra reveal dramatic differences compared to those from 2001/2002 presented by \cite{clark03}.
In our $K$ band spectrum, the strength of the Na {\sc i} doublet is significantly 
enhanced with respect to Br$\gamma$, and the He {\sc i} (2.112$\mu$m) line has all but disappeared. In the $H$ band, 
the changes are even more pronounced as the H {\sc i} series has transitioned from emission to absorption.
\cite{clark03} derived a temperature of 12000 K from their 2001/2002 spectra of G024.73+00.69. However, the
appearance of our spectra, in particular the absence of He {\sc i} lines, now implies a significantly lower temperature.   
This spectral variability, in addition to the photometric variability discovered by \cite{clark03}, further 
solidifies the confirmation of G024.73+00.69 as a bona fide LBV.

The new LBV candidates identified in the previous section are strikingly similar in appearance to these known 
sources. 
Star \#24 looks almost identical to \#45 (G026.47+00.02). 
Star \#14 resembles \#29 (WRAY 17-96) with overall weaker features and star \#17 is virtually a twin of 
\#44 (G 024.73+00.69). Stars \#46 and \#23a stand out due to their pronounced
[Fe {\sc ii}] and Fe {\sc ii} emission, and bear a strong resemblance to the spectrum of WRA 751 in the $K$ band (see 
Figure 8 of \citealt{morris96}).

Not surprisingly, our MIPS 24$\mu$m images of the shells (see Figure~\ref{f-charts}) reveal
significantly more detailed structure than the ISO images. The morphologies of the shells \#29 and \#44 are remarkably 
similar, showing dense shells with fairly even surface brightness.  
Both shells are also strongly detected at 8$\mu$m.
\cite{clark03} reported the possible detection of a much larger outer shell for \#44. The reality of this 
outer shell is difficult to confirm in our 8$\mu$m and 24$\mu$m images, given the complex structure in the general
background emission evident at both wavelengths. It is equally likely that the areas interpreted as ``lobes''
by \cite{clark03} are associated with unrelated embedded sources visible in the vicinity of that emission.
In contrast to shells \#29 and \#44, shell \#45 displays a more clumpy and filamentary structure. Some faint 
extended emission enhancements
are seen at 8$\mu$m coincident with the brighter areas at 24$\mu$m, but the morphology of the extended 
emission close to the central star is quite different from that at 24$\mu$m.

In addition to the known LBV / LBV candidates discussed so far, shell \#33 of our sample
is associated with the LBV candidate HD 316285. We did not obtain any spectra of the central source as part of this 
work, but
\cite{hillier98} present optical and near-IR spectra of the star. 
Their $H$ and $K$ band spectra look almost identical to those of our central source \#45. The star was known to be 
surrounded by a dusty shell based on the IRAS spectral energy
distribution \citep{mcgregor88}. The {\sl Spitzer} 24$\mu$m image of HD 316285 is the first observation to actually resolve 
the shell. Its morphology, as well as {\sl Spitzer} mid-IR spectroscopy of the shell is presented in \cite{morris08}.
Note that there is some extended emission at 8$\mu$m as well, however, the morphology of that 
structure is very different compared to that at 24$\mu$m. Overall, the characteristics of shell \#33 are quite similar 
to those of shell \#45. 
Amongst our newly identified LBV candidates (\#11a, 14, 17, 23a, 24, and 46), shells \#14, \#17, \#24, and \#46 most 
strongly resemble the shell morphologies of the known LBVs / LBV candidates, with shells \#11 and \#46 also exhibiting 
emission at 8$\mu$m.

\subsection{Optical Spectra~\label{s-opt}}

\begin{figure}[htb]
\epsscale{0.75}
\plotone{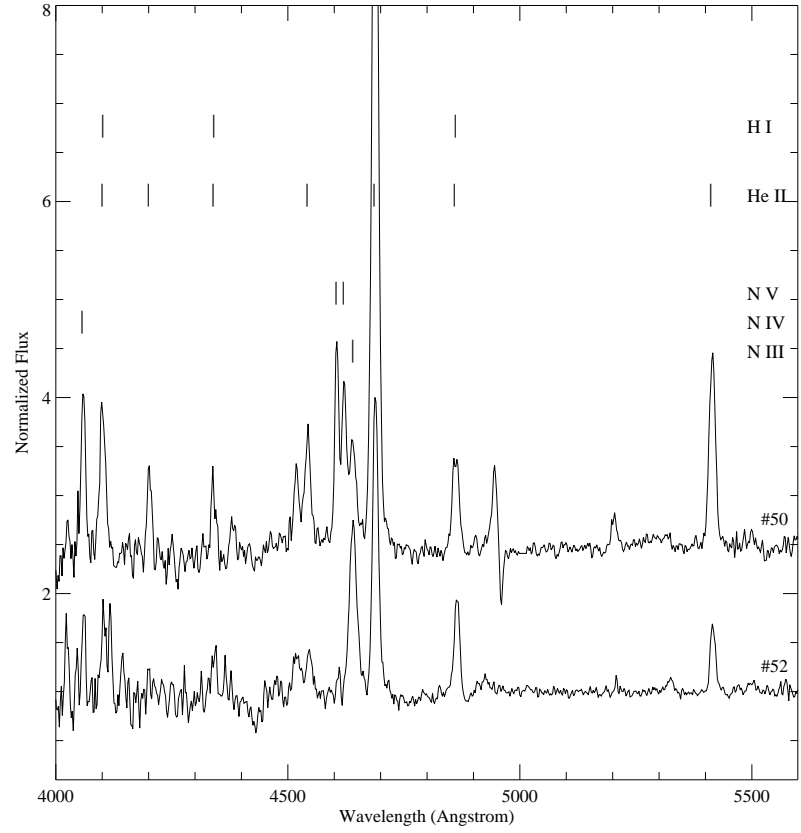}
\plotone{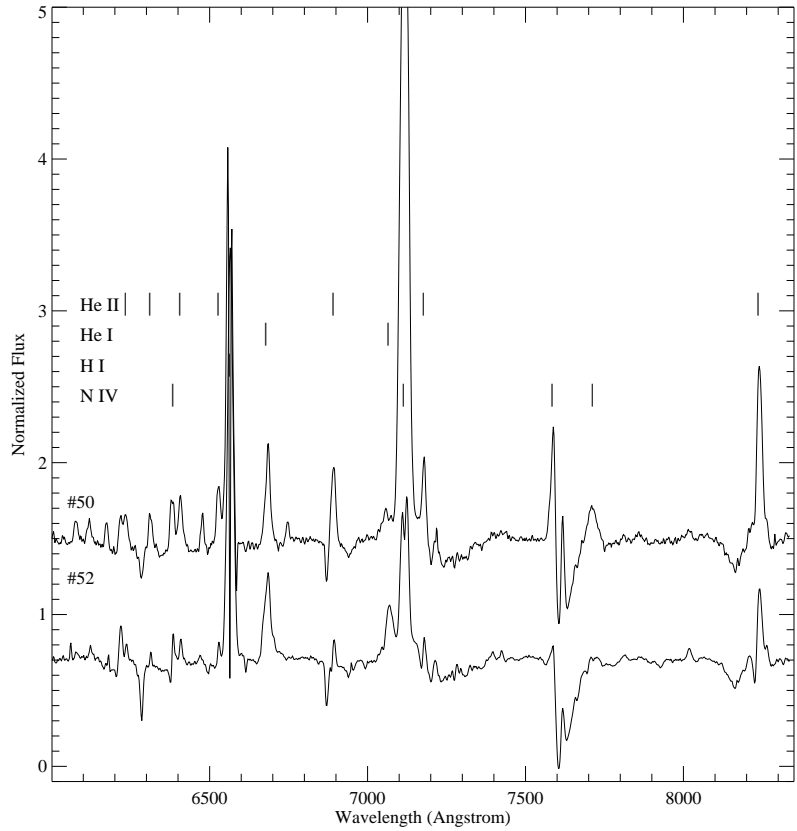}
\figcaption{The optical spectra of the newly discovered WN type WR stars \#50 (WN6o) and \#52 (WN7(h)). The spectra have been 
continuum normalized and offset for display purposes.  
\label{f-optwn} }
\end{figure}

Out of the nine central sources observed in the optical regime, two -- 
stars \#50 and \#52 -- are newly discovered WN stars (Figure~\ref{f-optwn}). We measured the equivalent 
width of N {\sc iv} $\lambda$4057, the N {\sc iii} $\lambda$4640 blend, and N {\sc v} $\lambda$4604, 4620 and followed the classification scheme 
outlined in \cite{conti90}. In addition, we compared our red spectra to those of the WR spectral 
catalog by \cite{vreux83}. We determine a spectral type of WN6 for source \#50.
Note that this source was previously classified as a planetary nebula, PN G029.0+00.4, its discovery 
dating back to 1966 \citep{abell66}. 
For star \#52, we derive a subtype 
of WN7, confirming the classification obtained earlier from our IR spectrum of this source. We also  
checked for the presence of hydrogen following the methodology of \cite{smith96}, and arrive at 
a classification of WN6o for star \#50 and WN7(h) for star \#52. 

\begin{figure}[htb]
\epsscale{0.75}
\plotone{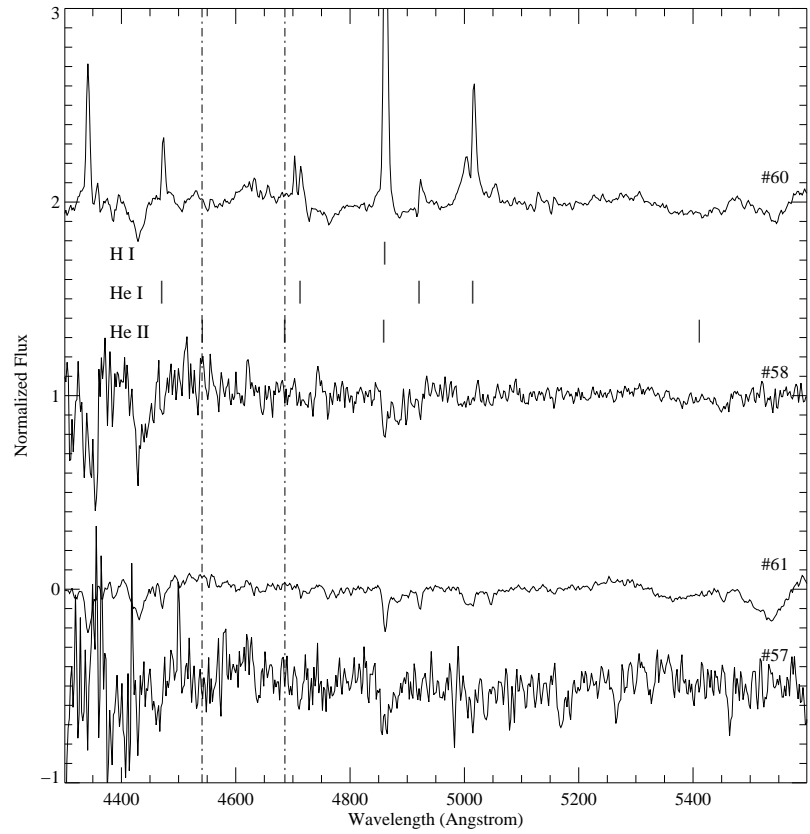}
\plotone{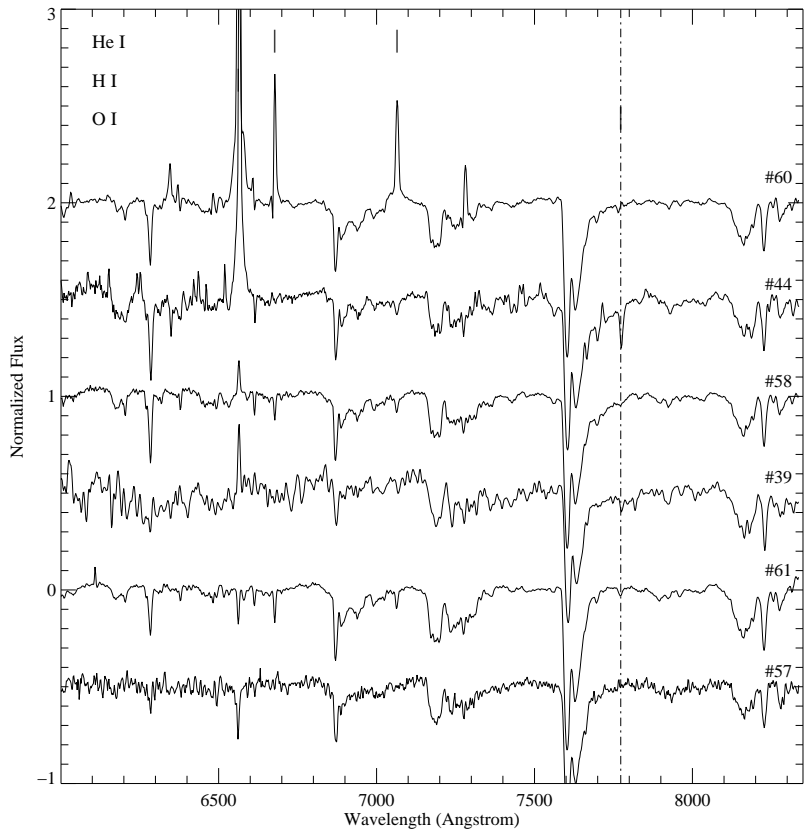}
\figcaption{The spectra of the sources classified as B stars (except for source \#57 which shows a F/G 
type spectrum). 
\label{f-optb} }
\end{figure}

We utilized the spectral atlases of \cite{allen95} and \cite{torres93} for an estimate
of the spectral type for the remainder of the central sources. We also checked these 
spectral types through comparison with observations
of spectral type standards obtained on the same nights as the targets.  
Source \#59b is an early M type star, $\sim$M2--M4 (the spectrum is not shown). 
The other six source spectra are displayed in Figure~\ref{f-optb}. In general, the blue spectra are 
quite noisy due to the highly reddened nature of these sources. Stars \#39 and \#44 were too faint 
to extract a signal at the shorter wavelengths, hence only red spectra could be obtained. 
Source \#58, \#60, and \#61 are 
classified as B type stars ($\sim$B0--B5) based on the presence of He {\sc i} (absorption or emission) 
and absence of He {\sc ii} $\lambda$4541 and 
$\lambda$4686 \citep{walborn90}. Source \#60 has a pure H {\sc i}, He {\sc i} emission spectrum.
Source \#44 is the known LBV G024.73+00.69 discussed in the previous section.
Source \#39 most likely also belongs to the group of B stars, but the 
red spectrum is quite noisy and it is difficult to gauge the reality of weak absorption features. 
Star \#57 most closely
matches an F/G type spectrum ($\sim$F5--G5). We are unable to determine the luminosity class for these 
sources based on our 
spectra. Note that the modulation in the continuum of the blue spectra (starting around 5300\AA) is an 
artifact of the data reduction process.   
 
\section{Infrared Photometry}

\begin{figure}[htb]
\epsscale{1.2}
\plotone{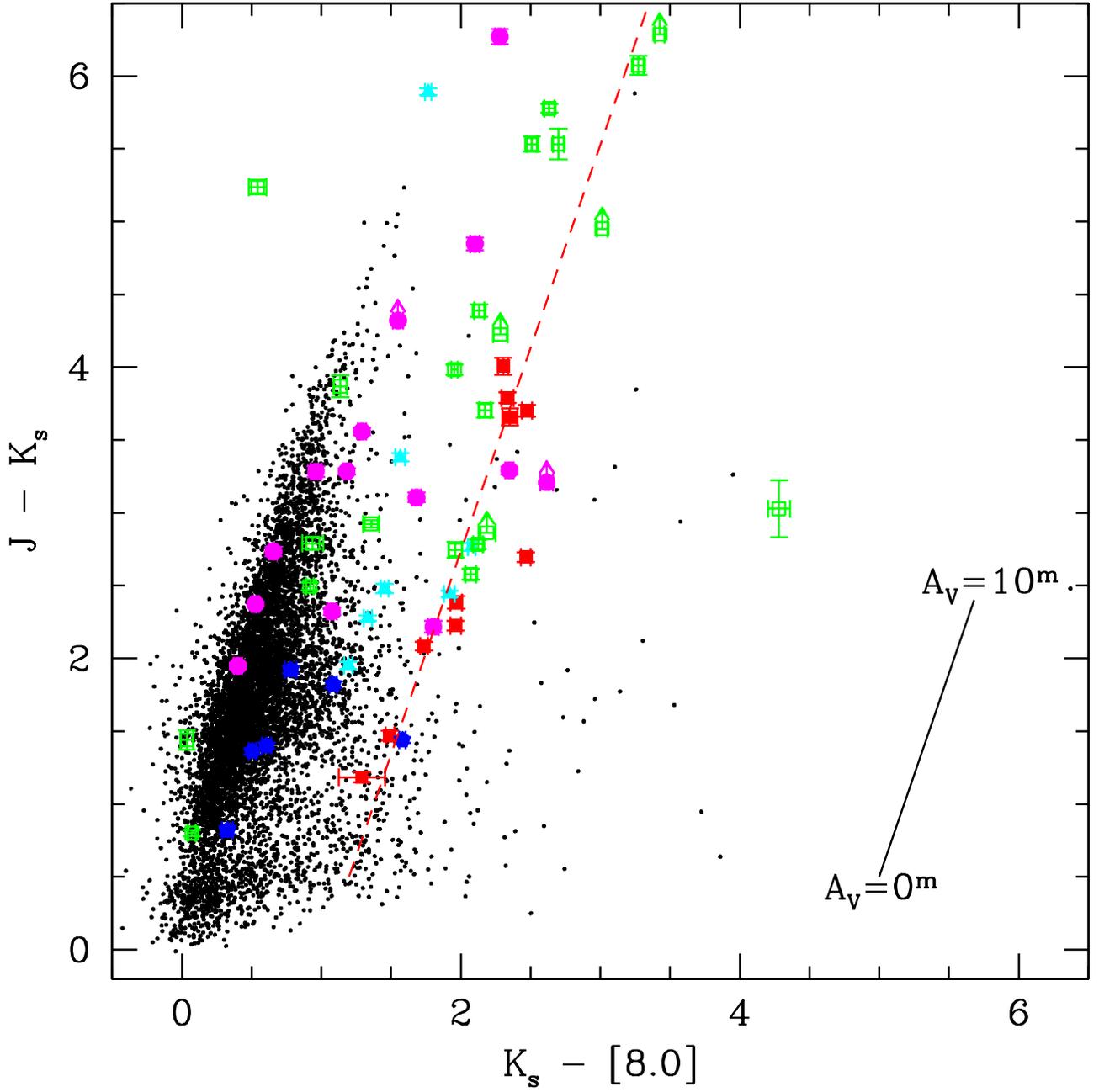}
\figcaption{Color--color diagram for the sources in our sample. WR stars are indicated in red, 
LBVs / LBV candidates in cyan, B stars in blue, and F-M stars in magenta.  
Green symbols indicate the central sources that are listed in 
Table~\ref{t-targets}, which have not yet been observed spectroscopically. 
The comparison population of sources (black points) is composed 
of the photometrically most reliable 
point sources of a representative  
$1\arcdeg \times\ 1\arcdeg$ ``slice'' of the Galactic plane from the GLIMPSE survey. 
The reddening vector was adapted from \cite{indebetouw05}.
\label{f-cc} }
\end{figure}

2MASS and GLIMPSE photometry for each source has been assembled in Table~\ref{t-shellphot}.
Following \cite{mauerhan09} and \cite{hadfield07} we constructed a color-color diagram (Figure~\ref{f-cc}) 
to explore any correlations between the known observed, and still 
unobserved sources of our sample.  
The comparison population of sources (black points) is composed of the photometrically most reliable 
point sources of a representative  
$1\arcdeg \times\ 1\arcdeg$ ``slice'' of the Galactic plane from the GLIMPSE survey. The various types of sources
we have identified spectroscopically are indicated by different colors (see the caption of Figure~\ref{f-cc} 
for details).
 
Given the relatively small sample size for each group, one might expect
limited insight from the distribution of points in color space, however, some general trends seem to emerge.  
As already noted in \cite{mauerhan09} and \cite{hadfield07}, the WR stars (red points) separate quite cleanly from 
the main locus of field stars.  
The WR stars in our sample form a surprisingly well-defined and tight sequence, roughly parallel to 
the reddening vector and offset to redder $Ks -$[8.0] colors from the bulk of the distribution of 
``normal'' stars. The direction of the reddening vector at the locus of our WR stars is indicated by the dashed line. 
Compared to the analogous figure presented in 
\cite{mauerhan09}, about half of our WR points have larger $J-Ks$ values, indicating a more 
heavily reddened population. 
B stars and LBV candidates appear to populate the locus in 
between the normal stars and the WR track. Obviously, there is some overlap between the location of these 
different sources in the color-color diagram, since stars \#24, \#44, and \#60 lie on the sequence defined by the 
WRs in our sample.  
Note that there are also some outliers among the F-M group of sources that are on or close 
to the WR sequence (stars \#23b, 26b, and 57). For the late type stars, this shift to redder colors might indicate a mid-IR 
excess due to cool circumstellar material. These objects could constitute a significant source
of contamination for color--based searches for WR stars. 

Having identified the preferred loci for the various types of central stars, we now turn our attention 
to the as yet unclassified sources (green points). 
The reddest outlier at $Ks -$[8.0]$>4$ is source \#59a, the bluest at $H-Ks \approx 5.2$ is source \#27b. 
Their nature is unclear. Stars \#41, \#47, \#48
and \#54 are tightly clumped on the WR sequence at $Ks -$[8.0]$\approx 2.1$ and hence are the best 
candidates for WR stars waiting to be identified. Stars \#2, \#25, \#34 and \#55 are found among the 
main locus of field stars and most likely belong to the group of F-M type stars.  
The remainder of the sources of unknown type (with photometry in all bands) are sources 
\#12, 15, 19, 27a, 30, 35, 38, 41, 
47, 48, 49, 51, 53, 54 and 56. We expect a mix of WR, LBV candidates, and F-M type stars among this 
group, but are unable to make any more finely differentiated predictions. 

\section{Shell Morphology, SED, and Size}

Morphologically, our 24$\micron$ shells separate broadly into two categories: fainter, more elliptical shells with somewhat 
irregular rims, and strikingly circular nebulae with sharply defined edges. Shells associated with late type stars 
(our F-M types) overwhelmingly belong to the first category (with the exception of shells \#8 and \#57), while the early 
type stars are usually exhibiting the most circular shells (exceptions are \#16, 18, 25, and 32). No distinguishing 
characteristics are evident in the appearance of the shells between WR stars, LBV candidates and B stars. 
The morphology of some of the shells surrounding known LBV stars has already been discussed in Section~\ref{s-lbv}.
Based on the general morphological distinction 
at 24$\mu$m between late and early type stars, we conclude that the hot star is likely producing the shell in 
the case of the double central sources containing both a late and early type star (\#23, \#28). 
We also note that the morphology of shell \#52 provides evidence for multiple, discrete mass loss events. 
Several concentric rings can be discerned even in the presence of significant structure in the overall 
background emission.
 
As noted in Section~1, many of the shells detected at 24$\mu$m do not exhibit equivalent 
emission at 8$\mu$m: 37 out of 
the 61 shells listed in Table~\ref{t-targets} (60\%) would not have been identified as shell sources at 8$\mu$m. 
Considering each of the spectral type subcategories, both of the known PNe, five out of the nine late type stars, 
none of the B type stars, 
two out of the nine Oe/WN and five out of the 10 Be/B[e]/LBV type stars show 8$\mu$m shell emission.
The mid-IR spectroscopic observations of some LBV and WR shells imply that the non-detection of 8$\mu$m
emission could simply be explained by a reduction in flux density at this wavelength relative to that at 24$\mu$m.  
The spectra presented by \cite{voors00} and \cite{barniske08} show a steep increase in emission towards longer 
wavelength starting at around 10$\mu$m.
At the same time, we cannot exclude the possibility that emission only observed 
at 24$\mu$m might be caused by a strong line feature in the spectrum, such as the [OIV] line producing 
the shell discovered by \cite{morris06}.

Based on the general morphological properties of the as yet unclassified sources, we speculate that 
the shells \#2, 12, 25, 34, 59a are formed by late type stars. Recall that we predicted sources \#2, 25, 34, and 55 
to likely be stars of type F-M from their location in the color-color diagram, supporting this conclusion.
We predict that shells \#7, 27, 30, 35, 37, 38, 41, 47, 48, 49, 51, 54, and 56 are associated with early type
central stars.  
Shells \#15, 19, 40, 42, 53, 55 remain ambiguous, partially because none of these   
have a prominent central source at 24$\mu$m, unlike all of the other shells. 

The observed radii of the 24$\mu$m shells listed in Table~\ref{t-targets} range from 0.14--2.4\arcmin\ with an 
average value of $\sim 0.7$\arcmin.  
The physical sizes for the shells surrounding late type stars have been derived in Section~\ref{s-late} and are 
listed in Table~\ref{t-size}. We now also estimate the dimensions of the shells around WR type sources among our sample, 
adopting the intrinsic colors and absolute $Ks$ band magnitudes for the various WR subtypes presented by 
\cite{crowther06}.
Utilizing the extinction ratios for the $JHKs$ bands from \cite{indebetouw05}, we first calculate an average extinction $A_{Ks}$ 
based on the observed 2MASS $J-Ks$ and $H-Ks$ colors (provided in Table~\ref{t-shellphot}). This extinction is combined with 
the observed $Ks$ magnitude and the assumed $M_{Ks}$ for the appropriate WR subtype, and we thus arrive at the distances and 
physical shell sizes listed in Table~\ref{t-wrsize}. Note that in addition to our newly identified WR stars \#10, \#11b, 
\#50 and \#52, we also list the known WR stars \#9 \citep{mauerhan09} and \#16 \citep{shara09}. The shells around these 
two sources have not been noticed previously, so the physical size estimate for the shells are new calculations, while
the distances and $A_{Ks}$ 
values for these stars were already presented in their respective discovery references.  The WC9 star \#11b is one 
of two possible central sources for shell \#11 (the fitted center of the shell lies in between the positions
of the two stars; see Section~8). We classify the other star, \#11a, as belonging to the group of 
Be/B[e]/LBV type stars, hence either star could be producing the observed shell, although circumstellar nebulae are more common
to B[e]/LBV stars than the more highly evolved WC stars. The close association of two such 
rare sources makes it likely that they are located at the same distance and could indicate the 
presence of a small cluster of stars. An investigation of the surrounding sources might be worthwhile. In any case, 
a distance estimate for one of the sources will determine the physical size of the shell. Unfortunately, \cite{crowther06}
do not provide a $M_{Ks}$ for WC9 stars, only intrinsic IR colors. We have therefore adopted the $M_{Ks}$ magnitude 
of star E from their Table~9 as a rough estimate.

The extinction, distance and size scales listed in Table~\ref{t-wrsize} seem reasonable, except in the case of star \#50, 
where a large distance (d=16 kpc) is paired with a relatively low extinction value ($A_V=5.7$). This is one of the few objects 
that exhibits 
shell emission at optical wavelengths. As already mentioned, the shell was classified as a planetary nebula, but we were unable 
to find any previous classification of the central star (see Section~\ref{s-opt} and 8).  

Comparing the radii of our WR shells to those reported in the literature, our shells generally fit well within the 
observed range, possibly somewhat skewed to the smaller end of the distribution. 
An optical study of eight WR ring nebulae by \cite{gruendl00} finds radii between 2.1--9.0 pc, and \cite{chu91}
quotes a few to 30 pc as the typical diameters of WR shells. \cite{marston96} surveyed 156 WR stars with IRAS and finds 49 ``probable or suspected shells'' ranging in size between
5--150 pc in radius. Due to the low spatial resolution of IRAS, these detections are clearly biased towards the largest
shells.  

We are unable to determine the physical sizes for the remainder of our shells, mostly comprising B/Be/B[e]/LBV type
stars. It is problematic to derive accurate distances for field B supergiants because they lack
luminosity diagnostics in their spectra. Detailed modeling of the stellar IR spectra as well as the shells to constrain the 
luminosity and reddening of the stars and determine shell parameters such as mass and dust composition, are beyond the 
scope of this paper and will be addressed in future work. \cite{clark03} summarize the best estimates
for the radii of LBV and candidate LBV shell nebulae, which range from 0.1--3.6 pc, most of them smaller than $\sim 1$ pc.

\section{Summary and Discussion}

We have discovered a large number of prominent, highly symmetric 24$\mu$m shells surrounding  
bright central sources. 
We were able to determine the nature of the apparent central sources for 45 of the 62 shells, 
10 based on literature searches and 35 from dedicated spectroscopic follow-up. 
Our infrared and optical spectroscopy has revealed a mix of early and late type stars. 
However, the prevalence of rare, massive stars stands out. We find six bona fide WR stars in our 
sample (four of which are new identifications), as well as four stars which we tentatively classify as 
Oe/WN. Most strikingly, our sample of shells contains four known LBV type stars (one confirmed, three candidates), 
and we classify an additional six sources as LBV candidates based on the similarities between 
the spectra. We realize that the classification of a particular source as an LBV involves 
a host of criteria, including photometric and spectroscopic variability, that still need to be
investigated for these new sources. However, given the small numbers of known LBVs, the presence 
of a shell combined with the spectroscopic characteristics make this relatively large sample
of new candidates an exciting discovery. 
We also identify five new Be stars in our sample. The distinction between these and what we 
characterize as LBV candidates is somewhat arbitrary and simply conveys less peculiar spectral
properties. It is entirely possible that these stars harbor additional LBV candidates, 
especially considering LBV spectral variability and the fact that the LBVs AG Car and HR Car 
have been classified as Be stars. 

The detailed evolutionary sequence for the most massive stars is still a matter of intense debate. 
The standard model of massive star evolution (see, e.g., the review by \citealt{crowther07}) expounds 
that O stars with initial masses of 25--40$M_\odot$ 
will undergo an LBV and RSG
phase on their way to becoming a WR star. More massive stars on the other hand, are thought to avoid the RSG phase. 
\cite{smithowocki06} contend that substantial mass loss during an LBV stage is critical to the formation 
of WR stars. Analysis of the nebulae surrounding the known LBVs support 
the view that the nebulae formed during a BSG and not during a RSG phase \citep{lamers01}.  
\cite{smithconti08} strongly argue in favor of the notion that 
H-rich WN type stars precede the LBV stage, while H-poor WN stars represent a post-LBV evolutionary phase. 
In this context we point out that our Oe/WN classified sources all show strong H emission, and hence could
be interpreted as pre-LBV stars. In fact, we speculate that a large number of the central stars 
associated with our most well-defined, circular shells represent a narrow evolutionary phase, closely
related or identical to the LBV phase. In depth analysis of the shell emission is needed to further 
investigate this hypothesis. 
Finally, we predict that our upcoming observations of the remainder of our sample will reveal additional members 
of this pre-LBV/LBV group among the as yet unclassified shell central stars.

\section{Appendix: Notes on Individual Objects~\label{s-notes}}

\begin{description}
\item[1] IRAS 11419-6228 is coincident with the central source
position. 

\item[2] MO 1-184, classified as a OB- star by \cite{muzzio77}, 
is offset by 35.18\arcsec\ from our central source position and has a 
24$\mu$m counterpart at the edge of the shell, coincident with an emission enhancement in the shell.  
Hence MO 1-184 is distinct from the 
central source associated with shell \#2. 
Note that the 24$\mu$m central source
is offset from the IRAC 3.6 and 8.0$\mu$m source. In the 2MASS images, several
blended sources are visible at the central source position. The 2MASS source listed in Table~\ref{t-targets} 
corresponds to the source 
that coincides with the brightest IRAC source
within 1\arcsec\ of the center position.

\item[4] IRAS 14173-6124 is offset by 21\arcsec\ away from the central source, but is 
likely associated with 
the central source. At 2MASS wavelengths there is a group (cluster?)
of bright stars that coincides with the southeastern border of the shell. 

\item[5] IRAS 15039-5806 is 64\arcsec\ away and clearly associated with a different bright 
24$\mu$m point source. In 2MASS, several sources are blended at the central source position.  

\item[6] The fitted geometric center of the ellipse falls halfway between the two bright 
24$\mu$m sources. 

\item[8] IRAS 15293-5602 is offset by 32\arcsec\ from the central source and clearly 
associated with a bright 24$\mu$m source on the east 
rim of the shell. There are four bright 24$\mu$m sources within the shell, source \#8a is at the 
geometric center, but we also obtained a spectrum of the source immediately to the west (\#8b).  

\item[9] This WR star was identified based on 2MASS and GLIMPSE photometric selection 
criteria by \cite{mauerhan09}. The presence of the shell, however, has not been noticed previously. 

\item[10] IRAS 15421-5323.

\item[11] 2MASS J15491137-5508516 (= IRAS 15452-5459), 260\arcsec\ distant, is classified as a 
post AGB star (proto-PN) in SIMBAD. 
It is associated with an extremely bright 24$\mu$m point source in the MIPSGAL images, but not with our shell. 
\cite{sahai07} present HST imaging of the post-AGB star which clearly shows a 
double lobed structure. However at 24$\mu$m there is no large scale extended emission associated with that
source, instead only our shell stands out.  
The fitted geometric center of the circular shell falls halfway between the two 
24$\mu$m sources listed in Table~\ref{t-targets}.

\item[12] IRAS 15517-5334 is 40\arcsec\ offset from our shell central source, but lies within the shell. The
association is uncertain. 

\item[14] IRAS 16254-4739 (15\arcsec offset) clearly associated with our shell. The only references 
for this object are \cite{vanderWalt96} and \cite{bronfman96}, who  
search for 6.7 GHz methanol maser emission associated with IRAS sources selected to meet 
criteria for being ultracompact H {\sc ii} regions. No maser emission was detected for IRAS 16254-4739. 

\item[15] An association with IRAS 16278-4808 (28\arcsec\ away) is possible, but somewhat ambiguous.   
C* 2338, a carbon star with a position offset by 122\arcsec\, is clearly associated with another 24$\mu$m source. 

\item[16] \cite{shara09} list a WN5b star (Shara 1093$\_53$) near the center of this shell, 
which is difficult to pinpoint due to its diffuse outer edge. They did not detect the shell itself, and there 
are three bright sources within
about 8\arcsec\ of the central region. This is one case where selecting the source coincident with the 
brightest central 24$\mu$m source would have identified the wrong star, as the WN5 star is not associated with
any enhanced $24\mu$m emission. We note that IGR J16320-4751, an embedded high mass X-ray binary, is 
also found in the vicinity (157\arcsec\ offset).  


\item[18] IRAS 16396-4555, EM* VRMF 115, Hen 2-179, WRAY 16-232, SS73 63, has been classified as anything 
from PN to ``not a PN'', highly obscured Be star to M supergiant, or peculiar emission line star. A good summary
of the literature on this source, as well as an optical spectrum, are presented in 
\citet{pereira03}. \citet{pereira03} classifies the source as a reddened Be star. Note that 
the shell has a highly clumped morphology.  

\item[19] IRAS 16426-4504 (60\arcsec\ offset) is clearly associated with another 24$\mu$m source. IGR J16465-4507 
is located nearby (215\arcsec), but not coincident with our shell. The IGR source consists of a compact object and 
a blue supergiant companion of spectral type O9.5Ia \citep{nespoli08}. 

\item[21] The planetary nebula IC 4637 is discussed as a possible containing a physical double central source 
with a K4 companion at a separation of 2.42\arcsec\ by \citet{ciardullo99}.  
2MASS clearly shows two sources, but only one appears in the 2MASS Point Source Catalog. The GLIMPSE morphology of the PN shell is 
described in \citet{phillips08}.

\item[22] IRAS 17039-3952 is offset by 30\arcsec\, located halfway between our shell and another 
embedded source. The association is ambiguous.

\item[23] IRAS 17050-3921.

\item[26] HD 317949, classified as a K0 star in SIMBAD, is offset by 36\arcsec\, and definitely 
associated with another source at 24$\mu$m at the edge of the shell. 
The geometric center of the shell is located between the two 24$\mu$m sources. 

\item[27] \#27a is the source closest to the geometric center of the shell. At 2MASS wavelengths, 
the central two sources appear to be blended with additional sources.  

\item[28] The radio source GPA 357.18-0.14 \citep{langston00} is found at a separation of 24\arcsec. It is 
not associated with our central source, but located on the SW rim of 
the shell. Source \#28a is closest to the geometric center of the shell.  

\item[29] The LBV candidate WRAY 17-96 was originally thought to be a 
PN.  \citet{egan02} describe the 
detection of a ring nebula with MSX and show a $K$ band spectrum of the central source. 

\item[31] Planetary Nebula PN G003.5+02.7, discovered by \cite{boumis03} and independently by the MASH survey 
\citep{parker06}. Morphologically, the 24$\mu$m shell strongly resembles the H$\alpha$ shell from 
the MASH survey. Our identification of the central source is uncertain due to a slight 
offset of the selected star from the geometric center of the shell.   

\item[32] IRAS 17408-3027. \cite{raharto84} list [RHI84] 10-469 at the position of our central source and 
give a M2: type. No additional information is provided. 

\item[33] HD 316285 is a known LBV candidate. \cite{hillier98} present optical and near-IR spectra of the 
central source. The star is known to be surrounded by a dusty shell based on the IRAS spectral energy
distribution \citep{mcgregor88}. The MIPS 24$\mu$m image is the first observation to resolve the
shell. Note that there is some extended emission at 8$\mu$m as well, however, the morphology of that 
structure is very different from that at 24$\mu$m. 


\item[36] At 2MASS wavelengths, the central source is clearly a blend of two sources. However, 
only one source is listed in the 2MASS catalog. The location of the source is within the coverage area
of the MAGPIS survey \citep{helfand06}, however, no radio emission is detected from either the shell or
the central source.  


\item[41] The X-ray source XGPS-I J182833-102652 is located 118\arcsec\ away, but clearly associated 
with a different 24$\mu$m source and 
nebulosity. At 2MASS wavelengths, the central source is blended with at least one other star.
The location of the source is within the coverage area
of the MAGPIS survey, however, no radio emission is detected from either the shell or
the central source.  


\item[44] G 024.73+00.69 (= V481 Sct) is a known LBV \citep{clark05} with a known shell. A $K$ band 
spectrum is shown in 
\citet{clark03}. No radio emission is detected from either the shell or the central source with the MAGPIS survey.

\item[45] G 26.470+0.021 is a known LBV candidate \citep{clark05} with a known shell. The MAGPIS 20cm image shows
bright emission coincident with the central source, as well as some nebular structure.   

\item[47 and 48] Source \#47 and \#48 form a spectacular complex of two neighboring shells at 24$\mu$m. We 
treat each shell component as a separate entity, although the combined
structure is also reminiscent of a double lobed nebula. However, there is no obvious central source at the 
``waist'' of this potential bipolar emission. It is unclear whether the two shells are interacting with one another or 
simply are a chance superposition of two independent shells. Due to the large size of each shell, the identification of 
the central sources is uncertain. At radio wavelengths, the outline of shell \#48 is faintly visible in 
the MAGPIS 20cm image, as well as a bright point source near the southern edge of the shell. 

\item[49] Both the shell and the central source are visible in the MAGPIS 20cm image.  

\item[50] This source is identified as planetary nebula PN G029.0+00.4 in SIMBAD. Its discovery dates back to \cite{abell66} where it 
is listed as A48. \cite{condon98} include the object in their catalog of PN in the NRAO VLA sky survey, but note it as ``Confused by 
complex background on the NVSS image.'' We were unable to find any previous classification of the central star, e.g., no central star type
is listed in \cite{zuckerman86}. The shell, but not the central star, is a strong radio source and clearly defined in the MAGPIS image.   

\item[51] SIMBAD lists [BSM2002] 18431-0312 4, a cloud of unknown nature at a distance of 53\arcsec. This is also 
IRAS 18431-0312 which in the 24$\mu$m MIPSGAL image is clearly associated with a different object, distinct from our shell. 
No radio emission is detected from 
either the shell or the central source with the MAGPIS survey.


\item[56] IRAS 18588+0350 is offset by 43\arcsec\, the association is uncertain. No radio emission is detected from 
either the shell or the central source with the MAGPIS survey.
\item[59] Source \#59a is closest to the geometric center of the shell. No radio emission is detected from 
either the shell or the central source with the MAGPIS survey.

\item[60] This source is listed as IRAS 19425+2411 and HBHA 2203-01 in SIMBAD. The only reference \citep{kohoutek99} simply includes 
the central source
in a catalog of H$\alpha$ emission line stars. No additional information is provided.  

\item[62] The central star of this known PN, NGC 6842, is a blend of two sources at 2MASS wavelengths. 
\end{description}

\acknowledgments

This work is based in part on archival data obtained with the Spitzer Space Telescope, which is 
operated by the Jet Propulsion Laboratory, California Institute of Technology under a contract with
NASA. Support for this work was provided by an award issued by JPL/Caltech.
Based on observations obtained at the Hale Telescope,
Palomar Observatory, as a part a continuing collaboration between the
California Institute of Technology, NASA/JPL, and Cornell University.
This publication makes use of data products from the 2 Micron All Sky
Survey, which is a joint project of the University of Massachusetts
and the Infrared Processing and Analysis Center/California Institute
of Technology, funded by the National Aeronautics and Space
Administration and the National Science Foundation. It also utilized
NASA's Astrophysics Data System Abstract Service and the SIMBAD
database operated by CDS, Strasbourg, France.

\begin{deluxetable}{cclccl}
\tabletypesize{\footnotesize}
\tablecaption{Shell Data
\label{t-targets}}
\tablehead{\colhead{Num}& \colhead{2MASSJ Designation} & \colhead{radius}& \colhead{central src} & \colhead{shell at} & \colhead{Comments\tablenotemark{1}}\\
\colhead{} &\colhead{} & \colhead{(arcmin)}&\colhead{24/8/3.6/1.1}&\colhead{8$\mu$m?}&\colhead{}}
\startdata
 1 & 11441803-6245210 & 0.53 x 0.35 & y/y/y/y   & no  & IRAS 11419-6228; Oe/WN ({\bf IR}), this work\\
 2 & 13012188-6244131 & 0.94 x 1.01 & y/y/y/y   & yes & \\
 3 & 13104384-6317457 & 0.45        & y/y/y/y   & no  & Oe/WN ({\bf IR}), this work \\
 4 & 14211002-6138199 & 1.08 x 2.38 & y/y/y/y   & yes & IRAS 14173-6124; M6 III / K5 I ({\bf IR}), this work\\
 5 & 15075621-5818250 & 0.50 x 0.42 & y/y/y/y   & no  & M0 III / K0 I ({\bf IR}), this work \\
 6a& 15092899-5847430 & 0.49 x 0.80 & y/y/y/y   & yes & M6 III / K5 I ({\bf IR}), this work \\
 6b& 15092951-5847551 &             & y/y/y/y   &     & M5 III / K4 I ({\bf IR}), this work \\
 7 & 15255986-5704403 & 0.17        & y/y/y/y   & yes & \\
 8a& 15330809-5612200 & 0.66        & y/y/y/y   & no  & K1 III / G7 I ({\bf IR}),this work \\
 8b& 15330663-5612219 &             & y/y/y/y   &     & M5 III / K4 I ({\bf IR}),this work \\
 9 & 15352652-5604123 & 1.07        & y/y/y/y   & no  & WN7 \citep{mauerhan09} \\
10 & 15455914-5332325 & 0.70        & y/y/y/y   & no  & IRAS 15421-5323; WN9h ({\bf IR}), this work \\
11a& 15484207-5507422 & 1.57        & y/y/y/y   & no  & Be/B[e]/LBV ({\bf IR}), this work\\
11b& 15484210-5507542 &             &           &     & WC9 ({\bf IR}), this work \\
12 & 15553788-5343402 & 1.24 x 2.42 & y/y/y/y   & yes & IRAS 15517-5334:\\
13 & 15581378-5257513 & 0.85        & y/y/y/y   & no  & Oe/WN ({\bf IR}), this work \\
14 & 16290377-4746264 & 0.33        & y/y/y/y   & yes & IRAS 16254-4739; Be/B[e]/LBV ({\bf IR}), this work \\
15 & 16313781-4814553 & 0.31 x 0.28 & n/y/y/y   & no  & IRAS 16278-4808: \\
16 & 16321298-4750358 & 1.02        & n/y/y/y   & no  & WN5b \citep{shara09} \\
17 & 16364278-4656207 & 0.30        & y/y/y/y   & no  & Be/B[e]/LBV ({\bf IR}), this work \\
18 & 16431636-4600424 & 0.95        & y/y/y/y   & no  & IRAS 16396-4555, SS 73 63; Be \citep{pereira03}\\
19 & 16461734-4508478 & 0.18        & y/y/y/y   & no  & \\
20 & 16493770-4535592 & 1.48        & y/y/y/y   & no  & Be ({\bf IR}), this work \\
21 & 17051043-4053071 & 0.25        & n/n/y/y   & yes & PN G345.4+00.1, IC 4637, Hen 2-193, WRAY 15-1607 \\
22 & 17072333-3956504 & 0.43 x 0.71 & y/y/y/y   & yes & M1 I ({\bf IR}), this work \\
23a& 17082913-3925076 & 0.79 x 0.73 & y/y/y/y   & no  & IRAS 17050-3921; Be/B[e]/LBV ({\bf IR}), this work \\ 
23b& 17082930-3925158 &             & y/y/y/y   &     & M2 III / K2 I ({\bf IR}), this work \\
24 & 17110094-3945174 & 0.46        & y/y/y/y   & no  & Be/B[e]/LBV ({\bf IR}), this work \\ 
25 & 17352942-3246562 & 0.55 x 0.47 & y/y/y/y   & no  &  \\                 
26a& 17370371-3147467 & 1.09 x 0.92 & y/y/y/y   & yes & M0 III / K0 I ({\bf IR}), this work\\  
26b& 17370394-3147349 &             & y/y/y/y   &     & G9 III ({\bf IR}), this work\\
27a& 17374754-3137333 & 0.51        & y/y/y/y   & yes &  \\
27b& 17374730-3137370 &             & n/y/y/y   &     &  \\ 
28a& 17391899-3124239 & 0.55        & y/y/y/y   & no  & OB ({\bf IR}), this work\\ 
28b& 17391913-3124142 &             & y/y/y/y   &     & M6 III / K5 I ({\bf IR}), this work\\ 
29 & 17413543-3006389 & 1.0         & y/y/y/y   & yes & WRAY 17-96, Hen 3-1453, LBV candidate, ({\bf IR}) \\ 
30 & 17421401-2955360 & 0.28        & y/y/y/y   & yes &  \\ 
31 & 17433908-2431525 & 0.24        & n/y/y/y   & yes & PN G003.5+02.7\\
32 & 17435981-3028384 & 1.04 x 0.76 & y/y/y/y   & no  & star M2:; Be ({\bf IR}), this work\\
33 & 17481403-2800531 & 1.54 x 1.97 & y/y/y/y   & yes & HD 316285, Hen 3-1482, WRAY 15-1777, LBV candidate\\
34 & 17493539-2649302 & 0.68 x 0.79 & y/y/y/y   & yes & \\
35 & 18005762-2433467 & 0.17        & n/y/y/y   & yes & IRAS 17578-2433 \\
36 & 18022233-2238002 & 0.42        & y/y/y/y   & yes & Oe/WN ({\bf IR}), this work \\
37 & 18035667-2256000 & 0.21        & y/n/y/y   & yes &  \\
38 & 18070516-2015163 & 0.36        & y/y/y/y   & no  &  \\
39 & 18133121-1856431 & 0.28        & n/y/y/y   & no  & Be: ({\bf OPT}), this work  \\
40 & 18233420-1525076 & 0.27        & n/y/y/y   & no  & \\
41 & 18284156-1027056 & 0.77        & y/y/y/y   & no  & \\
42 & 18321739-0916139 & 0.14        & n/n/y/y   & no  & \\ 
43 & 18333954-0807084 & 0.55        & y/y/y/y   & no  & G7 I ({\bf IR}), this work\\
44 & 18335528-0658386 & 0.74        & y/y/y/y   & yes & G 024.73+00.69, V481 Sct, LBV, ({\bf IR, OPT})\\
45 & 18393224-0544204 & 1.39 x 1.04 & y/y/y/y   & yes & G 026.47+00.02, 2MASS J18393224-0544204, LBV candidate, ({\bf IR})\\
46 & 18415965-0515409 & 0.43        & y/y/y/y   & yes & Be/B[e]/LBV ({\bf IR}), this work \\
47 & 18420630-0348224 & 1.67        & y/y/y/y   & no  & \\
48 & 18420827-0351029 & 1.16        & y/y/y/y   & no  & \\
49 & 18422247-0504300 & 0.47        & y/y/y/y   & no  & IRAS 18397-0507 \\
50 & 18424692-0313172 & 0.29        & n/y/y/y   & yes & PN G029.0+00.4; WN6 ({\bf OPT}), this work\\
51 & 18455593-0308297 & 0.40        & y/y/y/y   & yes & \\
52 & 18492733-0104207 & 1.8         & y/y/y/y   & no  & WN7 ({\bf IR, OPT}), this work\\
53 & 18503980+0004453 & 0.33        & n/y/y/y   & no  & IRAS 18481+0001 \\
54 & 18510295-0058242 & 1.07        & y/y/y/y   & no  & \\
55 & 18530582+0011358 & 0.24        & n/y/y/y   & no  & \\
56 & 19011669+0355108 & 0.52        & y/y/y/y   & yes & IRAS 18588+0350:\\
57 & 19042098+0600001 & 0.74        & y/y/y/y   & no  & F/G ({\bf OPT}), this work \\
58 & 19325284+1742303 & 0.78        & y/y/y/y   & no  & B0e--B5e  ({\bf OPT}), this work\\
59a& 19385510+2127550 & 0.89 x 0.49 & y/y/y/y   & yes & \\
59b& 19385569+2127483 &             &           &     & M2--M4 ({\bf OPT}), this work\\
60 & 19443759+2419058 & 0.53        & y/y/y/y   & no  & IRAS 19425+2411, HBHA 2203-01; Be ({\bf OPT}, this work)\\
61 & 19444295+2311337 & 0.46        & y/y/y/y   & no  & B0--B5 ({\bf OPT}), this work\\
62a& 19550232+2917178 & 0.42        & n/ / /y   &\nodata& PN G065.9+00.5, NGC 6842, Hen 2-451  \\
62b& 19550249+2917198 &             & y/ / /y   &     &  
\enddata
\tablenotetext{1}{For detailed notes on each object see Section~8}
\end{deluxetable}

\begin{deluxetable}{clrrc}
\tablecaption{Late Type Sources 
\label{t-size}}
\tablewidth{3.5in}
\tablehead{\colhead{Source}& \colhead{Type} & \colhead{A$_V$} & \colhead{d} & \colhead{size} \\
\colhead{} & \colhead{} & \colhead{(mag)} & \colhead{(kpc)} & \colhead{(pc)}}
\startdata
4    &  M6 III & 29.8 & 1.1 & $0.34 \times 0.75$  \\
     &  K5 I   & 32.5 & 2.6 & $0.80 \times 1.77$  \\
5    &  M0 III & 7.8  & 1.2 & $0.17 \times 0.14$  \\
     &  K0 I   & 10.4 & 6.1 & $0.89 \times 0.75$  \\
6a   &  M6 III & 12.0 & 3.2 & $0.46 \times 0.75$  \\ 
     &  K5 I   & 14.7 & 7.6 & $1.08 \times 1.76$  \\
6b   &  M5 III & 13.8 & 2.5 & $0.36 \times 0.58$  \\
     &  K4 I   & 16.7 & 7.6 & $1.09 \times 1.77$  \\
8a   &  K1 III & 14.4 & 0.6 & 0.11         \\
     &  G7 I   & 15.3 & 9.6 & 1.85         \\
8b\tablenotemark{1}   &  M5 III & 12.2 & 4.1 & 0.78         \\
     &  K4 I   & 15.0 & 12.5& 2.39         \\
22   &  M1 I   & 8.6  & 4.3 & $0.53 \times 0.88$  \\
23b\tablenotemark{1}  &  M2 III & 13.1 & 2.7 & $0.63 \times 0.58$  \\
     &  K2 I   & 15.7 & 11.6& $2.67 \times 2.47$  \\
26a  &  M0 III & 10.2 & 2.1 & $0.67 \times 0.56$  \\
     &  K0 I   & 12.8 & 11.0& $3.50 \times 2.96$  \\
26b  &  G9 III & 9.6  & 1.4 & $0.44 \times 0.37$  \\
     &  G5 I   & 10.4 & 23.7& $7.51 \times 6.34$  \\
28b\tablenotemark{1}  &  M6 III & 18.2 & 5.3 & 0.86         \\
     &  K5 I   & 20.9 & 12.6& 2.01         \\
43   &  K1 III & 24.8 & 0.3 & 0.44         \\ 
     &  G7 I   & 25.6 & 4.6 & 0.74        
\enddata
\tablenotetext{1}{Probably not producing the shell, see text for details. }
\end{deluxetable}

\begin{deluxetable}{crrrrrrr}
\tabletypesize{\footnotesize}
\tablecaption{Shell Central Source Photometry 
\label{t-shellphot}}
\tablehead{\colhead{Num}& \colhead{2MASS J} & \colhead{2MASS H} & \colhead{2MASS Ks} & \colhead{3.6$\mu$m\tablenotemark{1}} & \colhead{4.5$\mu$m\tablenotemark{1}} & \colhead{5.8$\mu$m\tablenotemark{1}} & \colhead{8.0$\mu$m\tablenotemark{1}}}
\startdata
1 &  9.999(23) & 8.720(31) &  7.775(23)&{\it 6.982(26)} & \nodata       &  6.175(37)    &  5.813(30)   \\
2 & 15.090(66) &12.459(75) & 11.221(36)& 10.281(31)     & 10.250(42)    & 10.001(36)    & 10.085(33)   \\     
3 & 12.271(23) &11.004(20) & 10.186(19)&  9.516(31)     &  9.065(41)    &  8.824(28)    &  8.450(22)   \\ 
4 & 12.719(48) & 8.618(55) &  6.448(15)&{\it 5.151(112)}&  \nodata      &  4.288(29)    &  4.169(29)   \\ 
5 &  9.282(20) & 7.815(21) &  6.959(13)&{\it 6.600(61)} &{\it 6.144(57)}&  5.900(37)    &  5.882(33)   \\ 
6a& 10.061(19) & 7.905(39) &  6.776(7) &{\it 6.697(74)} &  \nodata      &  5.731(33)    &  5.594(26)   \\  
6b& 10.870(23) & 8.520(41) &  7.311(9) &{\it 6.617(46)} &{\it 6.482(46)}&  6.142(35)    &  6.020(27)   \\ 
7 & 15.759(94) &13.628(77) & 12.390(57)& 11.929(309)    & 11.851(194)   & 11.280(97)    &  \nodata     \\
8a& 11.678(28) & 9.688(20) &  8.573(19)&  7.572(36)     &  7.274(47)    &  7.043(32)    &  6.890(27)   \\
8b& 11.472(21) & 9.218(20) &  8.189(29)&  7.510(47)     &  7.592(46)    &  7.330(34)    &  7.228(23)   \\
9 & 13.841(35) &12.386(39) & 11.457(27)& 10.583(49)     & 10.120(59)    &  9.843(47)    &  9.490(31)   \\ 
10& 14.576(34) &12.217(36) & 10.874(21)&  9.447(64)     &  9.023(55)    &  8.620(48)    &  8.399(29)   \\ 
11a& 7.929(23) & 6.725(45) &  5.977(13)& \nodata        &{\it 5.142(68)}&  4.943(26)    &  4.783(27)   \\ 
11b&11.185(25) & 9.708(28) &  8.490(21)&{\it 7.203(145)}&{\it 6.632(42)}&  6.239(37)    &  6.024(28)   \\ 
12& 12.380(25) & 8.637(17) &  6.603(17)&{\it 5.215(182)}&  \nodata      &{\it 3.923(25)}&{\it 3.969(33)} \\
13& 13.774(25) &11.369(29) &  9.984(27)&  8.652(44)     &  8.185(39)    &  7.854(37)    &  7.650(34)   \\
14&  8.668(27) & 7.141(43) &  6.188(15)&{\it 6.749(106)}&{\it 5.626(63)}&  5.076(31)    &  4.735(23)   \\
15&\llap{$>$}16.567&12.595(25)& 10.281(18)& 8.173(46)   &  7.462(40)    &  7.107(30)    &  6.854(30)   \\
16& 14.998(52) &12.697(33) & 11.340(26)& 10.049(49)     &  9.514(44)    &  9.196(88)    &  8.988(46)   \\
17& 11.711(18) & 7.889(39) &  5.820(15)&{\it 4.915(41)} &{\it 4.519(34)}&  4.046(17)    &  4.054(15)   \\ 
18&  6.258(17) & 5.077(45) &  4.205(33)& \nodata        & \nodata       & \nodata       & \nodata      \\  
19& 15.380(102)&11.867(30) &  9.847(21)&  8.091(36)     &  7.562(41)    &  7.252(36)    &  7.148(32)   \\ 
20&  7.242(17) & 6.086(33) &  5.422(17)&  \nodata       &{\it 5.022(55)}&  4.518(32)    &  4.336(19)   \\
21& 11.364(61) &11.343(91) & 10.964(63)&  \nodata       &  \nodata      & \nodata       & \nodata      \\
22&  7.009(9)  & 5.325(13) &  4.635(13)&  \nodata       &  \nodata      &  4.096(28)    &  4.109(23)   \\
23a&10.492(21) & 8.219(51) &  7.111(19)&{\it 6.463(87)} &{\it 6.164(55)}&  5.714(26)    &  5.545(29)   \\ 
23b&12.081(21) &10.032(24) &  8.790(17)&  7.686(47)     &  7.151(53)    &  6.788(32)    &  6.442(25)   \\
24&  9.728(21) & 8.041(33) &  6.963(17)&{\it 6.673(154)}&{\it 6.075(88)}&  5.239(37)    &  4.884(23)   \\ 
25& 10.749(19) & 9.101(20) &  8.254(17)&  7.633(36)     &  7.443(29)    &  7.295(29)    &  7.336(19)   \\ 
26a&11.260(23) & 9.342(23) &  8.527(21)&  7.949(31)     &  8.093(29)    &  7.885(30)    &  7.872(19)   \\ 
26b&12.556(23) &11.014(26) & 10.339(32)&  9.365(49)     &  9.038(90)    &  8.711(34)    &  8.536(21)   \\ 
27a&11.771(34) & 9.248(38) &  7.788(9) &{\it 6.838(42)} &{\it 6.191(45)}&  5.957(28)    &  5.834(22)   \\
27b&13.701(6)  &10.588(41) &  8.464(49)&  8.095(99)     &{\it 8.316(89)}&  7.747(42)    &  7.921(38)   \\
28a& 8.440(19) & 7.586(29) &  7.078(11)&{\it 6.864(33)} &  6.657(44)    &  6.548(27)    &  6.573(23)   \\ 
28b&$>$12.898  &10.078(36) &  8.578(31)&  7.464(28)     &{\it 7.502(29)}&  7.080(28)    &  7.029(18)   \\ 
29&  6.707(13) & 5.520(33) &  4.796(11)& \nodata        & \nodata       & \nodata       &  \nodata     \\ 
30& 13.382(33) &10.540(33) &  8.993(29)&  7.677(39)     &  7.158(35)    &  6.843(27)    &  6.862(23)   \\ 
31& 13.822(51) &12.685(44) & 12.382(43)& 12.222(52)     & 12.265(76)    & 11.915(98)    & 12.346(134)  \\   
32&  5.743(13) & 4.781(75) &  4.070(33)& \nodata        & \nodata       & \nodata       & \nodata      \\
33&  4.817(33) & 4.227(75) &  3.712(234)& \nodata       & \nodata       & \nodata       & \nodata      \\ 
34&  8.906(13) & 8.319(39) &  8.106(19)&  8.026(29)     &  8.138(27)    &  8.076(23)    &  8.035(19)   \\ 
35& 12.714(15) &10.840(21) &  9.790(20)&  8.856(30)     &  8.494(28)    &  8.294(23)    &  8.432(53)   \\ 
36& 13.604(51) &11.073(32) &  9.597(27)&  8.365(45)     &  7.762(44)    &  7.499(30)    &  7.293(24)   \\ 
37& $>$14.022  &13.333(100)& 12.930(64)& 12.489(98)     & 12.227(186)   &  \nodata      & \nodata      \\ 
38& 14.867(42) &12.671(42) & 11.162(27)&  9.904(44)\tablenotemark{2} & \nodata  &  9.180(41)\tablenotemark{2} & \nodata  \\
  &  & &  & 9.913(26)\tablenotemark{3} &  9.433(39)\tablenotemark{3}  &  9.153(28)\tablenotemark{3} & 8.988(30)\tablenotemark{3} \\ 
39& 11.569(17) &10.340(25) &  9.648(19)&  9.256(76)     &  9.044(63)    &  8.896(47)    &  8.867(37)   \\ 
40& $>$15.565  &14.487(71) & 13.733(73)& 13.034(78)     & 13.001(124)   &      \nodata  &    \nodata   \\ 
41& $>$14.189  &12.630(74) & 11.327(39)& 10.090(144)    &  9.688(62)    &  9.395(57)    &  9.139(48)   \\ 
42& $>$13.148  &12.703(64) & 12.400(41)& 11.852(96)     & 11.583(106)   & 11.648(178)   &    \nodata   \\ 
43& 12.998(38) & 9.851(22) &  8.151(19)&  7.100(87)     &{\it 6.392(47)}&  6.108(35)    &  6.050(28)   \\ 
44&  8.363(17) & 6.836(35) &  5.920(15)&{\it 4.464(71)} &{\it 4.462(60)}&{\it 3.935(32)}&  4.001(35)   \\ 
45&  7.997(13) & 6.526(21) &  5.608(9) &{\it 4.796(81)} &{\it 4.211(63)}&{\it 3.811(29)}&  \nodata     \\ 
46&  7.960(5)  & 6.534(13) &  5.684(19)&     \nodata    &    \nodata    &  4.608(27)    &  4.351(21)   \\ 
47& 11.946(26) &10.216(20) &  9.162(19)&  8.277(37)     &  7.630(42)    &  7.430(28)    &  7.039(30)   \\ 
48& 11.845(28) &10.260(29) &  9.267(24)&  8.425(38)     &  7.876(49)    &  7.667(32)    &  7.196(28)   \\ 
49& $>$13.635  &11.101(35) &  8.684(23)&{\it 6.873(66)} &   \nodata     &  5.750(30)    &  5.672(31)   \\
50& 13.508(24) &12.834(27) & 12.325(26)& 11.693(64)     & 11.245(102)   & 11.063(85)    & 11.035(164)  \\
51& 15.447(61) &11.505(21) &  9.373(23)&  7.389(45)     &  6.717(47)    &  6.316(35)    &  6.100(28)   \\ 
52& 10.943(25) &10.068(32) &  9.473(23)&  8.850(52)     &  8.455(37)    &  8.297(41)    &  7.981(23)   \\ 
53& $>$16.395  &13.645(57) & 12.169(35)& 10.855(59)     & 10.433(56)    & 10.163(65)    &  9.886(39)   \\ 
54& 11.875(37) &10.141(46) &  9.130(38)&  8.146(53)     &  7.721(54)    &  7.481(35)    &  7.169(26)   \\
55& 14.428(29) &12.538(25) & 11.638(24)& 10.923(57)     & 10.652(69)    & 10.597(77)    & 10.700(72)   \\ 
56& 15.225(49) &11.661(22) &  9.693(19)&  8.080(36)     &  7.644(47)    &  7.248(33)    &  7.184(32)   \\ 
57& $>$14.783  &12.960(38) & 11.574(30)& 10.204(76)     &  9.551(54)    &  9.270(45)    &  8.958(35)   \\ 
58& 10.207(17) & 9.306(14) &  8.806(14)&  8.478(39)     &  8.379(47)    &  8.234(40)    &  8.199(29)   \\ 
59a&16.489(183)&14.667(140)& 13.461(71)& 11.825(54)     & 11.204(62)    & 10.624(60)    &  9.180(33)   \\
59b& 9.860(21) & 8.387(31) &  7.912(5) &  7.535(60)     &  7.800(44)    &  7.547(32)    &  7.511(30)   \\ 
60&  8.857(13) & 8.016(19) &  7.418(15)&{\it 6.734(30)} &{\it 6.421(38)}& 6.176(22)     &  5.837(16)   \\
61&  9.916(15) & 9.375(27) &  9.096(23)&  8.863(33)     &  8.889(48)    &  8.728(31)    &  8.769(32)   \\
62a&15.681(111)&$>$14.462  & $>$14.007 & \nodata        & \nodata       & \nodata       & \nodata\\ 
62b&15.557(66) &$>$14.253  & $>$13.680 & \nodata        & \nodata       & \nodata       & \nodata \\
\enddata
\tablenotetext{1}{All magnitudes from the GLIMPSE I/II/3D survey. Data in italics are from the less reliable archive deliveries.}
\tablenotetext{2}{GLIMPSE I.}
\tablenotetext{3}{GLIMPSE II.}
\end{deluxetable}

\begin{deluxetable}{clrrrrc}
\tablecaption{WR Distances and Shell Sizes 
\label{t-wrsize}}
\tablewidth{4.5in}
\tablehead{\colhead{Source}& \colhead{Type} & \colhead{$M_{Ks}$} & \colhead{$A_{Ks}$} & \colhead{$A_{V}$} &\colhead{d} & \colhead{size} \\
\colhead{} & \colhead{} & \colhead{(mag)} & \colhead{(mag)} & \colhead{(mag)} &  \colhead{(kpc)} & \colhead{(pc)}}
\startdata
 9   & WN7  & $ -4.77$  &   1.27  & 11.1  & 9.8 &  3.04 \\
 10  & WN9h & $ -5.92$  &   2.32  & 20.3  & 7.8 &  1.60 \\
 11b\tablenotemark{1} & WC9  &  $-6.10$\tablenotemark{2}   &   1.70  & 14.9  & 3.8 &  1.73 \\
 16  & WN5b & $ -4.77$  &   2.09  & 18.3  & 6.4 &  1.89 \\
 50  & WN6  & $ -4.41$  &   0.65  & 5.7  & 16.5 &  1.39 \\
 52  & WN7(h) & $ -4.77$&   0.66  & 5.8  &  5.2 &  2.72 \\
\enddata
\tablenotetext{1}{Possibly not producing the shell, see text for details. }
\tablenotetext{2}{No $M_{Ks}$ is given for WC9 stars in Table A1 of \cite{crowther06}. We adopt the 
absolute magnitude of star E from their Table 9.} 
\end{deluxetable}

\end{document}